\pdfoutput=1
\documentclass[extra]{gji}

\usepackage[dvipsinames]{xcolor}

\usepackage[%
            colorlinks=True,%
            urlcolor=blue,%
            linkcolor=blue,%
            citecolor= blue,%
            pdfauthor={T. Gastine},%
            pdftitle={pizza: an open-source spectral code for 
quasi-geostrophic convection},]{hyperref}
\usepackage{timet}
\usepackage{lineno}
\usepackage{xcolor}
\usepackage{amsmath,amsfonts,amssymb}
\usepackage{graphicx}
\usepackage{journals}
\usepackage{booktabs}

\urldef\ggg\url{%
https://agupubs.onlinelibrary.wiley.com/action/downloadSupplement?doi=10.1002%2F
2016GC006438&file=ggge21074-sup-0002-2016GC006438-s02.zip}

\DeclareMathOperator{\arcsinh}{arcsinh}

\title{\texttt{pizza}: an open-source pseudo-spectral code for 
spherical quasi-geostrophic convection}
\author[Thomas Gastine]
  {Thomas Gastine$^1$ \\
  $^1$ Institut de Physique du Globe de Paris, Sorbonne Paris Cit\'e,
Universit\'e Paris-Diderot, UMR 7154 CNRS, 1 rue Jussieu, F-75005 Paris,
France}
\date{Received \today; in original form \today}
\pagerange{\pageref{firstpage}--\pageref{lastpage}}
\volume{200}
\pubyear{2018}

\def\vec#1{\ensuremath{\mathchoice{\mbox{\boldmath$\displaystyle#1$}}
{\mbox{\boldmath$\textstyle#1$}}
{\mbox{\boldmath$\scriptstyle#1$}}
{\mbox{\boldmath$\scriptscriptstyle#1$}}}}

\begin{document}

\label{firstpage}

\maketitle

\begin{summary}

We present a new pseudo-spectral open-source code nicknamed \texttt{pizza}. It 
is dedicated to the study of rapidly-rotating Boussinesq convection under the 
2-D spherical quasi-geostrophic approximation, a physical hypothesis 
that is appropriate to model the turbulent convection that develops in 
planetary interiors.
The code uses  a Fourier decomposition in the azimuthal direction and 
supports both a Chebyshev collocation method and a sparse Chebyshev integration 
formulation in the cylindrically-radial direction. It supports several temporal
discretisation schemes encompassing multi-step time steppers as well as 
diagonally-implicit Runge-Kutta schemes.
The code has been tested and validated by comparing 
weakly-nonlinear convection with the eigenmodes from a linear solver. 
The comparison of the 
two radial discretisation schemes has revealed the superiority of the Chebyshev 
integration method over the classical collocation approach both in terms of 
memory requirements and operation counts.
The good parallelisation efficiency enables the computation of large problem 
sizes with $\mathcal{O}(10^4\times 10^4)$ grid points using several thousands 
of ranks. This allows the computation of numerical models in 
the turbulent regime of quasi-geostrophic 
convection characterised by large Reynolds $Re$ and yet small Rossby numbers 
$Ro$. A preliminary result obtained for a strongly supercritical numerical 
model with a small Ekman number of $10^{-9}$ and a Prandtl number of unity 
yields $Re\simeq 10^5$ and $Ro \simeq 10^{-4}$. \texttt{pizza} is hence an 
efficient tool to study spherical quasi-geostrophic convection in a parameter 
regime inaccessible to current global 3-D spherical shell models.
\end{summary}


\section{Introduction}

Convection under rapid rotation is ubiquitous in astrophysical bodies.
The liquid iron cores of terrestrial planets or the atmospheres of the gas 
giants are selected examples where turbulent convection is strongly influenced 
by rotational effects \citep[e.g.][]{Aurnou15}. Such turbulent flows are 
characterised by very large 
Reynolds numbers $Re > 10^8$ and yet small Rossby numbers $Ro < 10^{-5}$, $Ro$ 
being defined as the ratio between the rotation period and the convective 
overturn time. This specific combination of $Re \gg 1$ and $Ro \ll 1$ 
corresponds to the so-called \emph{turbulent quasi-geostrophic 
regime} of rotating convection \citep[e.g.][]{Julien12a,Stellmach14}. 
This 
implies that, in absence of a magnetic field, the pressure gradients balance the 
Coriolis force at leading order. As a consequence, the convective flow shows a 
pronounced invariance along the axis of rotation. At onset of rotating 
convection for instance, the flow pattern takes the 
form of quasi-geostrophic elongated columnar structures that have a typical 
size of $E^{1/3}$, 
where $E=\nu/\Omega d^2$ is the Ekman number with $\nu$ the kinematic 
viscosity, $\Omega$ the rotation frequency and $d$ the thickness of the 
convective layer \citep[e.g.][]{Busse70,Dormy04}. Convection in natural objects 
corresponds to extremely small Ekman numbers with for instance $E\simeq 
10^{-15}$ in the Earth core or $E\simeq 10^{-18}$ in the gas giants.
The quasi-geostrophy of the convective flow is expected to hold as long 
as the dynamics is dominated by rotation, or in other words as long as the 
buoyancy force remains relatively small compared to the Coriolis force 
\citep{Gilman77,Julien12a,King13,Cheng15,Horn15,Gastine16}.

Many laboratory experiments of rotating convection in spherical geometry have 
been carried out, either under micro-gravity conditions 
\citep[e.g.][]{Hart86,Egbers03}; or on the ground using the centrifugal force 
as a surrogate of the radial distribution of buoyancy 
\citep[e.g.][]{Busse74,Sumita03,Shew05}. Because of their limited size, those
experiments could only reach  $E \simeq 5\times 
10^{-6}$, far from the geophysical/astrophysical regime. In complement to the 
laboratory experiments, rotating convection in spherical geometry can also be 
studied by means of three-dimensional global
numerical simulations. Because of computational limitations, those numerical 
models are currently limited to $ E \gtrsim 10^{-7}$, $Re \lesssim 10^4$ and 
$Ro \gtrsim 10^{-3}$, hardly scratching into the turbulent quasi-geostrophic 
(hereafter QG) regime \citep{Gastine16,Schaeffer17}. Reaching lower Ekman 
numbers is hence mandatory to further explore this regime with $Re\gg 1$ and 
$Ro \ll 1$.

A way to alleviate the computational constraints inherent in global 3-D 
computations is to consider a spherical QG approximation of the 
convective flow 
\citep[e.g.][]{Busse86,Cardin94,Plaut02,Aubert03,Morin04,Gillet06,Calkins12,
Teed12,Guervilly17,More18} . The underlying assumption of the spherical QG 
approximation is that the leading-order cylindrically-radial 
and azimuthal velocity components are invariant along the axis 
of rotation $z$. 
Under this approximation, the variations of the axial 
vorticity along the rotation axis are also neglected and an averaging of the 
continuity equation along the rotation axis implies a linear dependence of 
the axial velocity on $z$ \citep{Schaeffer05,Gillet06}. The spherical QG 
approximation hence restricts
the computation of the evolution of the convective velocity to two dimensions 
only. This is a limitation compared to the 3-D QG convective models 
developed by \cite{Calkins13} which allow spatial modulations of the convective 
features along the rotation axis. Because of the radial distribution of the 
buoyancy forcing in spherical geometry, the temperature is not necessarily 
well-described by the 
quasi-geostrophic approximation. Spherical QG models with either 
a three-dimensional or a two-dimensional treatment of the temperature however 
yield very similar results \citep{Guervilly16}. Despite those 
approximations, the different implementations of the 2-D spherical QG models 
\citep[e.g.][]{Aubert03,Gillet06,Calkins12,Teed12,Guervilly17} have been found 
to compare favourably to 3-D direct numerical simulations in spherical geometry
\citep[e.g.][]{Aubert03,Schaeffer05,Plaut08}. This indicates that such 2-D 
spherical QG models could be efficiently used to explore the turbulent QG 
regime of convection with $E < 10^{-8}$ and $Re \gtrsim 10^5$, a parameter 
regime currently inaccessible to 3-D computations.
Quasi-geostrophy is expected to hold as long as the dynamics is 
dominated by rotation, or in other words as long as the buoyancy force remains 
relatively small compared to the Coriolis force 
\citep{Gilman77,Julien12a,King13,Cheng15,Horn15,Gastine16}.

The spatial discretisation strategy adopted in spherical QG models 
usually relies on a hybrid scheme with a truncated 
Fourier expansion in the azimuthal direction $\phi$ and second-order finite 
differences in the cylindrically-radial direction $s$ 
\citep[e.g.][]{Aubert03,Calkins12}. Note that \cite{Brummell93} and 
\cite{Teed12} rather employed a spectral Chebyshev collocation technique in $s$ 
but in the case of a cartesian QG model. The vast majority of those numerical 
codes adopt a pseudo-spectral approach where the nonlinear terms are treated in 
the physical space and time-advanced with an explicit Adams–Bashforth time 
scheme, while the linear terms are time-advanced in the Fourier space using a 
Crank–Nicolson scheme.
In contrast to 3-D models where several codes with active on-going 
developments are freely accessible to the community \citep[see][]{Matsui16}, 
there is a no open-source code for spherical QG convection available to the 
community.

The purpose of this study is precisely to introduce a new open-source 
pseudo-spectral spherical QG code, nicknamed \texttt{pizza}.  \texttt{pizza} is 
available at \url{https://github.com/magic-sph/pizza} as a free software that 
can be used, modified, and redistributed under the terms of the GNU GPL v3 
license.  The package also comes with a suite of \texttt{python} classes to 
allow a full analysis of the outputs and diagnostics produced by the code 
during its execution. The code, written in Fortran, uses a Fourier 
decomposition in $\phi$ and either a Chebyshev collocation or a sparse 
Chebyshev integration method  in $s$ 
\citep[e.g.][]{Stellmach08,Muite10,Marti16}. It supports a broad variety of 
implicit-explicit time schemes encompassing multi-step methods 
\citep[e.g.][]{Ascher95} and implicit Runge-Kutta schemes 
\citep[e.g.][]{Ascher97}. The parallelisation strategy relies on the Message 
Passing Interface (\texttt{MPI}) library.

The paper is organised as follows. Section~\ref{sec:model} presents the 
equations for spherical QG convection. Section~\ref{sec:rschemes} and 
\ref{sec:tschemes} are dedicated to the spatial 
and temporal discretisation schemes implemented in \texttt{pizza}. The 
parallelisation strategy is described in section~\ref{sec:mpi}. The code 
validation and several examples are discussed in section~\ref{sec:results} 
before concluding in section~\ref{sec:conclusion}.

\section{A quasi-geostrophic model of convection}

\label{sec:model}

Because  of the strong axial invariance of the flow under rapid rotation, the 
QG models approximate 3-D convection in spherical geometry by a 2-D
fluid domain which corresponds to the equatorial plane of a spherical shell.
Using the cylindrical coordinates $(s,\phi,z)$, the QG fluid domain 
hence corresponds to an annulus of inner radius $s_i$ and outer radius 
$s_o$ rotating against the $z$-axis with an angular frequency $\Omega$.
In the following, we adopt a dimensionless formulation of the spherical QG 
equations using the annulus gap $d=s_o-s_i$ as a reference 
length scale and the viscous diffusion time $d^2/\nu$ as the reference time 
scale. The temperature contrast $\Delta T$ between both boundaries defines the 
temperature scale. Gravity is assumed to grow linearly with the cylindrical 
radius $s$ and is non-dimensionalised using its value at the 
external radius $g_o$.

The formulation of the QG model implemented in \texttt{pizza} is based on the 
spherical QG approximation introduced by \cite{Busse86} and further expanded by 
\cite{Aubert03} and \cite{Gillet06} to include the effects of Ekman pumping.
Following \cite{Schaeffer05} and \cite{Gillet06} the axial velocity $u_z$ is 
assumed to vary linearly with $z$.
Under this assumption, the Boussinesq continuity equation under the spherical 
QG approximation yields
\begin{equation}
 \dfrac{1}{s}\dfrac{\partial (s u_s)}{\partial s}+\dfrac{1}{s}\dfrac{\partial 
u_\phi}{\partial \phi}+\beta u_s = 0\,,
 \label{eq:cont}
\end{equation}
where
\begin{equation}
 \beta = \dfrac{1}{h}\dfrac{\mathrm{d} h}{\mathrm{d} s} =-\dfrac{s}{h^2}\,,
\end{equation}
and $h=(s_o^2-s^2)^{1/2}$ is half the height of the geostrophic cylinder
at the cylindrical radius $s$. We adopt a vorticity-streamfunction formulation 
to fulfill the QG continuity equation~(\ref{eq:cont}). The cylindrically-radial 
and azimuthal velocity components are hence expanded as follows
\begin{equation}
 u_s = \dfrac{1}{s}\dfrac{\partial \psi}{\partial \phi},\quad
 u_\phi = \overline{u_\phi}-\dfrac{\partial \psi}{\partial s}-\beta \psi,
 \label{eq:vel_def}
\end{equation}
where the streamfunction $\psi$ accounts for the non-axisymmetric motions, 
while $\overline{u_\phi}$ corresponds to the axisymmetric zonal flow component, 
the overbar denoting an azimuthal average. The axial vorticity $\omega$ is then 
expressed by 
\begin{equation}
 \omega = \dfrac{1}{s}\dfrac{\partial(s\overline{u_\phi})}{\partial 
s}-\mathcal{L}_\beta \psi,
 \label{eq:psi}
\end{equation}
where the operator $\mathcal{L}_\beta$ is defined by
\[
 \mathcal{L}_\beta \psi = \Delta \psi+\dfrac{1}{s}\dfrac{\partial 
(\beta s \psi)}{\partial s}\,.
\]
In the above equation, $\Delta$ is the Laplacian operator in cylindrical 
coordinates. Under the QG approximation, the time evolution of the axial 
vorticity becomes

\begin{equation}
 \dfrac{\partial \omega}{\partial t} + \vec{\nabla}\cdot\left( \vec{u}\,\omega 
\right) = \dfrac{2}{E}\beta u_s - \dfrac{Ra}{Pr} \dfrac{1}{s_o}\dfrac{\partial 
\vartheta}{\partial \phi} +\mathcal{F}(E,\vec{u},\omega)+  \Delta \omega\,,
\label{eq:vort}
\end{equation}
where $\vartheta$ denotes the temperature perturbation.
The reader is referred to \cite{Gillet06} for a comprehensive derivation of 
this equation. In the above equation, $\mathcal{F}(E,\vec{u},\omega)$ 
corresponds to the Ekman-pumping contribution \citep{Schaeffer05} to 
non-axisymmetric motions expressed by 

\begin{equation}
\mathcal{F}(E,\vec{u},\omega) = -\Upsilon\left[
 \omega-\dfrac{\beta}{2}u_\phi+\beta\left(\dfrac{\partial}{\partial 
\phi}-\dfrac{5 s_o}{2h}\right) u_s\right]\,.
 \label{eq:pumping_full}
\end{equation}
where
\[
 \Upsilon = \left(\dfrac{s_o}{E}\right)^{1/2}\dfrac{1}{(s_o^2-s^2)^{3/4}}\,.
\]
To ensure a correct force balance in the azimuthal direction, the 
axial vorticity equation (\ref{eq:vort}) is supplemented by an equation 
dedicated to the axisymmetric motions \citep{Plaut02}. Taking a $\phi$-average 
of the azimuthal component of the Navier-Stokes equations yields
\begin{equation}
 \dfrac{\partial \overline{u_\phi}}{\partial t}+\overline{u_s\omega} = 
-\Upsilon\,\overline{u_\phi}+\Delta \overline{u_\phi} - 
\dfrac{\overline{u_\phi}}{s^2}\,,
\label{eq:uphi}
\end{equation}
where the first term in the right-hand-side corresponds to the Ekman-pumping 
contribution for the axisymmetric motions \citep{Aubert03}.
The governing equations for the temperature perturbation under the QG 
approximation is given by
\begin{equation}
 \dfrac{\partial \vartheta}{\partial t} + \vec{\nabla}\cdot \left(
\vec{u}\,\vartheta\right)+\beta u_s \vartheta + u_s\dfrac{\mathrm{d} 
T_c}{\mathrm{d} s} = \dfrac{1}{Pr}\Delta \vartheta\,,
\label{eq:temp}
\end{equation}
where $T_c$ is the conducting background state \citep{Aubert03,Gillet06}. In 
the case of a fixed-temperature contrast between $s_i$ and $s_o$, 
$T_c$ is given by
\[
 T_c = \dfrac{\alpha}{\ln{\eta}}\ln [(1-\eta)s],\quad 
\dfrac{\mathrm{d} T_c}{\mathrm d s} =\dfrac{\alpha}{s\ln\eta}\,,
\]
where $\alpha$ is a constant coefficient that can be used to rescale the 
temperature contrast to get a better agreement with the $z$-average of the 
conducting temperature of a 3-D spherical shell \citep{Aubert03,Gillet06}. In 
the case of fixed temperature boundary conditions,
\[
 \alpha = 
\dfrac{\eta}{1-\eta}\left\lbrace\dfrac{1}{(1-\eta^2)^{1/2}}\arcsinh\left[\dfrac{
(1-\eta^2)^{1/2}}{\eta}\right]-1\right\rbrace\,.
\]
The dimensionless equations (\ref{eq:psi}-\ref{eq:temp}) are governed by the 
Ekman number $E$, the Rayleigh number $Ra$ and the Prandtl number $Pr$ defined 
by
\begin{equation}
 E = \dfrac{\nu}{\Omega d^2},\quad Ra = \dfrac{\alpha_T g_o \Delta T 
d^3}{\nu\kappa}, \quad Pr= \dfrac{\nu}{\kappa}\,,
 \label{eq:controls}
\end{equation}
where $\alpha_T$ is the thermal expansion coefficient and $\kappa$ is the 
thermal diffusivity. 

We assume in the following no-slip and fixed temperature at both boundaries. 
This yields
\begin{equation}
 u_s =u_\phi = \vartheta = 0 \quad\text{at}\quad 
s=s_i,s_o\,.
\label{eq:bcs}
\end{equation}
With the definition of the streamfunction (Eq.~\ref{eq:vel_def}), this 
corresponds to
\begin{equation}
 \psi =\dfrac{\partial \psi}{\partial s} = \vartheta = \overline{u_\phi} = 0
 \quad\text{at}\quad s=s_i,s_o\,.
 \label{eq:bcs_psi_intro}
\end{equation}

\section{Spatial discretisation}

\label{sec:rschemes}

The unknowns $u_s$, $u_\phi$, $\omega$ and $\vartheta$ are 
expanded in truncated Fourier series in the azimuthal direction up to a maximum 
order $N_m$. For each field $f=[u_s,u_\phi,\omega,\vartheta]$, one has
\[
 f(s,\phi_k,t) \approx \sum_{m=-N_m}^{N_m} f_m(s,t)\,e^{\mathrm { i } m\phi_k 
}\, ,
\]
where $\phi_k =2\pi (k-1)/N_\phi$  with $k=1, ..., N_\phi$ defines
$N_\phi$ equally-spaced discrete azimuthal grid points. Since all the physical 
quantities are real, $f_{-m}^*=f_m$, where the star denotes a complex 
conjugate. Complex to real Fast Fourier Transforms (FFTs) can hence be employed 
to transform each quantity from a spectral representation to a grid 
representation
\begin{equation}
f(s,\phi_k,t)=2\,\sideset{}{'}\sum_{m=0}^{N_m}\Re\left\lbrace f_m(s,t)
\,e^{\mathrm{ i } m\phi_k } \right\rbrace\,,
\label{eq:fft}
\end{equation}
where the prime on the summation indicates that the $m=0$ coefficient needs to 
be multiplied by one half. The inverse transforms are handled by real to 
complex FFTs defined by
\begin{equation}
 f_m(s,t) = \dfrac{1}{N_\phi}\sum_{k=1}^{
N_\phi} f(s,\phi_k,t)\,e^{-\mathrm{i}m\phi_k}\,.
 \label{eq:ifft}
\end{equation}
Using $N_\phi \geq 3N_m$ prevents aliasing errors when treating the 
non-linear terms \citep{Orszag71,Boyd01}. This implies to discard the Fourier 
modes with $N_m<m\leq N_\phi$ when doing the direct FFT (\ref{eq:fft}) and to 
pad with zeroes when computing the inverse transforms (\ref{eq:ifft}).

In the radial direction, the Fourier coefficients $f_m$ are further expanded 
in truncated Chebyshev series up to degree $N_c-1$

\begin{equation}
 f_m(s_k,t) = C \,\sideset{}{''}\sum_{n=0}^{N_c-1}\widehat{f}_{mn
} (t)\,T_{n}(x_k)\,,
\label{eq:cheb}
\end{equation}
where the hat symbols are employed in the following to denote the Chebyshev 
coefficients. The discrete Chebyshev transform from a spectral representation 
to a grid representation is given by
\begin{equation}
 \widehat{f}_{mn}(t) = C\,\sideset{}{''}\sum_{k=1}^{N_r} f_m(s_k,t)
\,T_{n}(x_k)\,.
\label{eq:icheb}
\end{equation}
In the above equations $C=[2/(N_r-1)]^{1/2}$ is a normalisation factor and the 
double primes on the summations now indicate that both the first and the last 
indices are multiplied by one half. $T_n(x_k)$ is the $n$th-order 
first-kind Chebyshev polynomial defined by
\[
 T_n(x_k) = T_{kn} =  \cos[n\arccos(x_k)]= \cos\left[
\dfrac{\pi n (k-1)}{N_r-1}\right]\,,
\]
where
\[
 x_k = \cos\left[ \dfrac{\pi(k-1)}{N_r-1}\right], \quad k=1, ..., N_r,
\]
is the $k$th-point of a Gauss-Lobatto grid with $N_r$ collocation grid points.
For an annulus of inner radius $s_i$ and outer radius $s_o$, the Gauss-Lobatto 
interval that ranges from $-1$ to $1$ is remapped to the interval $[s_i,s_o]$ 
by the following affine mapping
\[
 s_k = \dfrac{s_o-s_i}{2}\,x_k+\dfrac{s_o+s_i}{2},\quad k=1, ..., N_r\,.
\]
The choice of using Gauss-Lobatto grid points also ensures that fast 
Discrete Cosine Transforms of first kind (DCTs) can be employed to compute the 
transforms between Chebyshev representation and radial grid space 
(\ref{eq:cheb}-\ref{eq:icheb}). \texttt{pizza} relies on the 
\texttt{FFTW}\footnote{\url{http://fftw.org/}}  library \citep{Frigo05} for 
all the FFTs and DCTs. This ensure that each single spectral transform
is computed in $\mathcal{O}(N\ln N)$ operations, where $N=[N_r,N_m]$.

\subsection{Spectral equations using Chebyshev collocation}

Several approaches can be employed to approximate the solution of a 
differential equation using Chebyshev polynomials. The most straightforward 
choice when dealing with a set of non-constant partial differential equations 
such as Eqs.~(\ref{eq:psi}-\ref{eq:temp}) is to resort to a Chebyshev 
collocation method \citep[e.g.][]{CHQZ}.
In this kind of approach, the unknowns can be either the 
Chebyshev coefficients $\widehat{f}_n$ or the values of the approximate 
solution at the collocation points $f(x_k)$. Both collocation techniques yield 
dense matrices with similar condition numbers \citep{Peyret02}. The first 
one has been widely adopted by the astrophysical and geophysical communities 
after the seminal work by \cite{Glatzmaier84}.

\subsubsection{Semi-discrete formulation}

Expanding $\omega$, $\psi$ and $\vartheta$ in Fourier and Chebyshev modes 
yield the following set of coupled semi-discrete equations for the time 
evolution of $\widehat{\omega}_m$ and $\widehat{\psi}_m$ for the 
non-axisymmetric modes 
with $m>0$	

\begin{equation}
  \begin{aligned}
 C\sideset{}{''}\sum_{n=0}^{N_c-1}\left\lbrace
 \left[\dfrac{\mathrm{d}}{\mathrm{d} t}T_{kn} -\mathcal{A}^C_{mkn} \right]
\widehat{\omega}_{mn}(t) 
+\mathcal{B}^C_{mkn}\widehat{\psi}_{mn}(t)\right\rbrace &	= \\
-\left[\dfrac{Ra}{Pr}\dfrac{\mathrm{i}m}{s_o}\right]\vartheta_{m}(s_k,t)- 
{\mathcal{N}_\omega}_m(s_k,t)\,&\, \\
C \sideset{}{''}\sum_{n=0}^{N_c-1} \left\lbrace 
T_{kn}\,\widehat{\omega}_{mn}(t)+\mathcal{C}^C_{mkn} 
\widehat{\psi}_{mn}(t)\right\rbrace& 
=0\,,
\end{aligned}
\label{eq:psiomcoll}
\end{equation}
where the collocation matrices are expressed by
\[
  \begin{aligned}
 \mathcal{A}^C_{mkn} =&
T_{kn}''+\dfrac{1}{s_k}T_{kn}'-\left[\dfrac{m^2}{s_k^2}+\Upsilon_k\right]T_{
kn} , \\
   \mathcal{B}^C_{mkn} 
= &\dfrac{\Upsilon_k\beta_k}{2}\,T_{kn}'+\\& \beta_k\left[\dfrac{
\beta_k\Upsilon
_k}{2}+\dfrac{\mathrm{i}m}{s_k}\left(\mathrm{i}m\Upsilon
_k-\dfrac{5s_o \Upsilon_k} { 2h_k } -\dfrac { 2 } { E } \right)\right]T_{kn} ,\\
  \mathcal{C}^C_{mkn} = & T_{kn}''+\left[\beta_k+\dfrac{1}{s_k}
\right]T_ 
{kn}'-\left[\dfrac{\mathrm{d}\beta_k}{\mathrm{d}s}+\dfrac{\beta_k}{s}+\dfrac { 
m^2 } { s_k^2 }\right]T_{kn}\,,
\end{aligned}
\]
In the above equations, the superscripts $^C$ have been introduced to 
differentiate the collocation matrices from the forthcoming sparse formulation. 
For clarity, a given function $f$ discretised at the collocation point 
$x_k$ is expressed as $f_k=f(x_k)$. $T'_{kn}$ and $T''_{kn}$ are the first and 
second derivative of the $n$th-order Chebyshev polynomial at the collocation 
point $x_k$. ${\mathcal{N}_\omega}_m(s_k,t)$ corresponds to the Fourier 
transform (\ref{eq:ifft}) of the advection terms that enters Eq.~(\ref{eq:vort})
\[
 {\mathcal{N}_\omega}_m(s_k,t) = \dfrac{1}{N_\phi}\sum_{j=1}^{N_\phi} 
\left[\vec{\nabla}\cdot(\vec{u}\,\omega)\right]e^{-\mathrm{i}m\phi_j}\,.
\] 
where $N_\phi=3N_m$ to ensure that the nonlinear terms are alias-free in $\phi$ 
\citep{Orszag71}. 

Instead of introducing the intermediate variable $\omega$, we 
could rather have substituted its definition (\ref{eq:psi}) into 
Eq.~(\ref{eq:vort}) to derive a single time-evolution equation that would 
depend on $\psi$ only. This would imply to solve an equation of the form
\[
 \dfrac{\partial}{\partial t}\left(\dfrac{\partial^2 \psi}{\partial 
s^2}\right)+\cdots = \dfrac{\partial^4 \psi}{\partial s^4}+\cdots
\]
Though appealing this strategy is however not viable since this kind of 
time-dependent problem has been shown to be unconditionally unstable when 
using Chebyshev collocation discretisation \citep{Gottlieb77,Hollerbach00}.

We proceed the same way to discretise the equations for the mean azimuthal flow 
$\overline{u_\phi}$ (\ref{eq:uphi})

\begin{equation}
  \begin{aligned}
  C\sideset{}{''}\sum_{n=0}^{N_c-1}&\left[
\dfrac{\mathrm{d}}{\mathrm{d} t}T_{kn} 
-T_{kn}''-\dfrac{1}{s_k}T_{kn}'+\right. \\
&\left.\left(\Upsilon_k+\dfrac{1}{s_k^2}
\right)
T_{kn}\right]\widehat{{{u_\phi}}_0}_{n}(t) =
-\mathcal{N}_{u_{\phi}}(s_k,t),
\end{aligned}
\label{eq:uphicoll}
\end{equation}
where the nonlinear term is expressed by
\[
 \mathcal{N}_{u_{\phi}}(s_k,t) =\dfrac{E}{2}\Upsilon_k 
u_{\phi_0}\omega_0+2\sum_{1}^{N_m} 
\Re\left\lbrace{u_s}_m\omega^*_m\right\rbrace\,.
\]
The first term in the right hand side corresponds to the self-interaction 
of the zonal wind \citep{Aubert03}. Finally, the spatial discretisation of the 
temperature equation (\ref{eq:temp}) yields
\begin{equation}
  \begin{aligned} 
C\sideset{}{''}\sum_{n=0}^{N_c-1}\left[
\dfrac{\mathrm{d}}{\mathrm{d} t}T_{kn} 
-\dfrac{1}{Pr}\left(T_{kn}''+\dfrac{1}{s_k}T_{kn}'-\dfrac{m^2}{s_k^2}
T_{kn}\right)\right ] \widehat{\vartheta}_{mn}(t) = \\
    \left[\dfrac{\mathrm i m}{s_k}\dfrac{\mathrm{d} T_c}{\mathrm{d} 
s}\right]\psi_{m}(s_k,t) - {\mathcal{N}_\vartheta}_m(s_k,t)\,,
 \end{aligned}
 \label{eq:tempcoll}
\end{equation}
where ${\mathcal{N}_\vartheta}_m(s_k,t)$ corresponds to the FFT 
of the nonlinear terms that enter Eq.~(\ref{eq:temp}):
\[
 {\mathcal{N}_\vartheta}_m(s_k,t) = \dfrac{1}{N_\phi}\sum_{j=1}^{N_\phi} 
\left[\vec{\nabla}\cdot(\vec{u} \vartheta) +\beta_k 
u_s\vartheta\right]e^{-\mathrm{i}m\phi_j}\,.
\]

\subsubsection{Boundary conditions}

In the collocation method, equations (\ref{eq:psiomcoll}), 
(\ref{eq:uphicoll}) and (\ref{eq:tempcoll}) are prescribed for the $N_r-2$ 
 internal collocation grid points. The remaining boundary points $s=s_i$ and 
$s=s_o$ are used to impose the boundary conditions (\ref{eq:bcs_psi_intro}). 
This implies that the singularity of $\beta$ and its derivatives at the outer 
boundary $s_o$ is not necessarily an issue when using the collocation method 
since boundary conditions provide additional constraints there. 
When a given physical field $f=[\psi,\omega,\vartheta,\overline{u_\phi}]$ is 
subject to Dirichlet boundary conditions at both boundaries, the following 
conditions on the Chebyshev coefficients $\widehat{f}_n$ should be fulfilled 
\citep[e.g.][Eq.~3.3.19]{CHQZ}
\begin{equation}
  \sideset{}{''}\sum_{n=0}^{N_c-1} \widehat{f}_{nm} = 0,\ s=s_o;\quad
  \sideset{}{''}\sum_{n=0}^{N_c-1} (-1)^{n} \widehat{f}_{nm} = 0,\ 
s=s_i\,,
\label{eq:bcs_coll_dirichlet}
\end{equation}
while for Neumann boundary conditions \citep[e.g.][Eq.~3.3.23]{CHQZ}
\begin{equation}
  \sideset{}{''}\sum_{n=0}^{N_c-1} n^2 \widehat{f}_{nm} = 0,\  
s=s_o;\   \sideset{}{''}\sum_{n=0}^{N_c-1} (-1)^{n+1} n^2 \widehat{f}_{nm} = 0, 
\ s=s_i\,.
\label{eq:bcs_coll_neumann}
\end{equation}

\begin{figure}
 \centering
 \includegraphics[width=8.4cm]{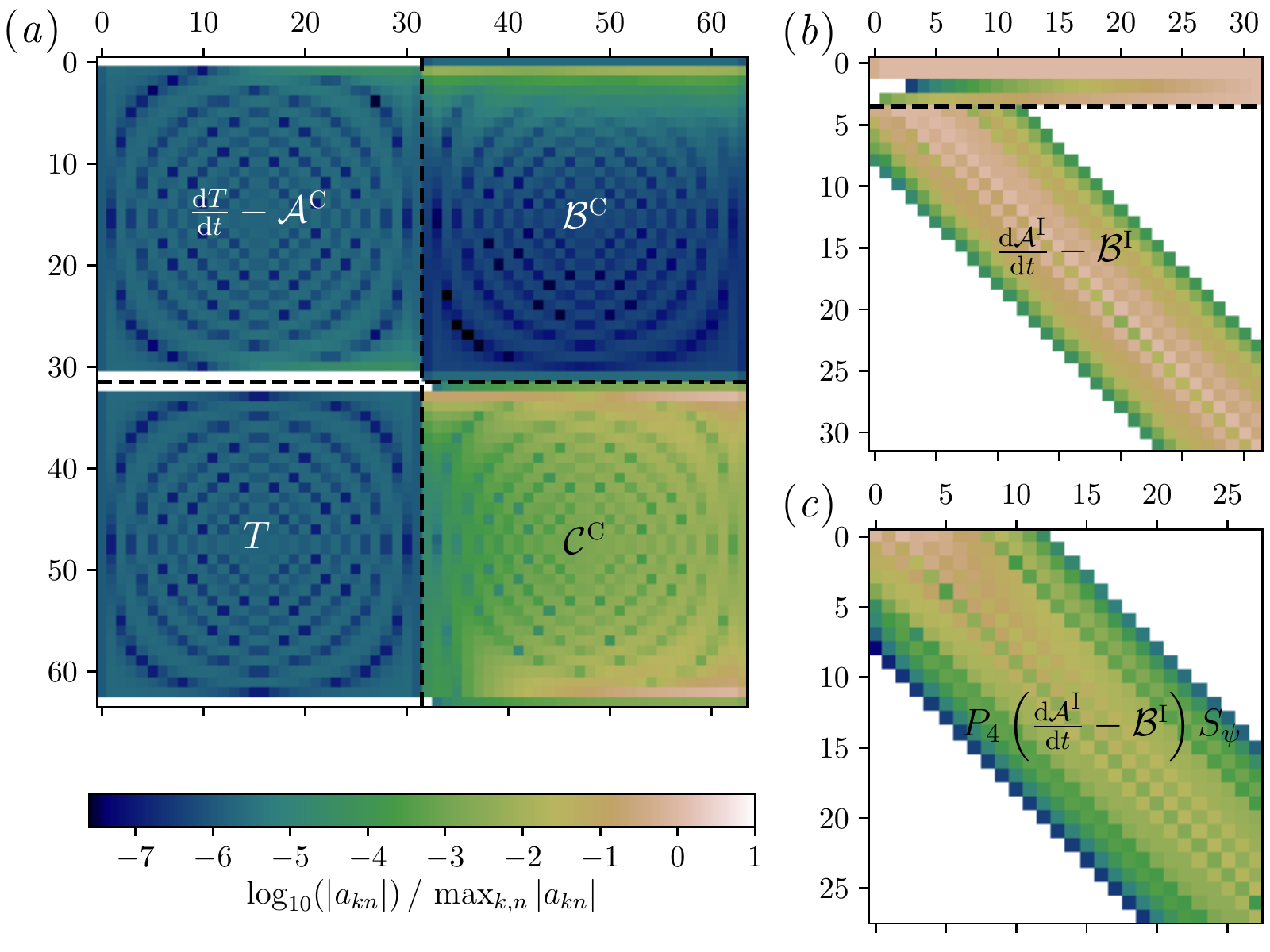}
 \caption{Representation of the coefficients of the left-hand-side matrices  
obtained for $m=4$ for a setup with $E=10^{-3}$, $Ra=3\times 10^4$ and $Pr=1$ 
and a CNAB2 time scheme with a fixed $\delta t = 10^{-4}$. (\textit{a}) 
corresponds to the collocation method (Eq.~\ref{eq:psiomcoll}).  $T$ corresponds 
to the matrix with the coefficients $T_{kn}=T_n(x_k)$. (\textit{b}) corresponds 
to the Chebyshev integration method with boundary conditions imposed as the 
first four tau lines (Eq.~\ref{eq:psiint}). (\textit{c}) corresponds to 
Chebyshev integration method with boundary conditions
enforced via a Galerkin formulation (Eq.~\ref{eq:psiint_galerkin}). For the 
three panels, the matrix coefficients have been normalised by their maxima such 
that they share the same color axis. Zero entries are displayed in white.}
 \label{fig:mat}
\end{figure}

Independently of the subsequent details of the chosen implicit-explicit time 
scheme employed to time advance the QG equations, Eq.~(\ref{eq:psiomcoll}) forms 
a complex-type dense matrix operator of size $(2N_r\times 2N_r)$ for each 
Fourier 
mode $m$. Figure~\ref{fig:mat}a shows the structure of the matrix that enters  
the left-hand-side of Eq.~(\ref{eq:psiomcoll}). The top $N_r$ rows corresponds 
to the time-dependent vorticity 
equation (\ref{eq:vort}), while the bottom $N_r$ rows corresponds to the 
streamfunction equation (\ref{eq:psi}). The four mechanical boundary conditions 
(\ref{eq:bcs_psi_intro}) are imposed on the first and last rows of the top-right 
and bottom-right quadrants of this matrix.


From a numerical implementation standpoint, Chebyshev polynomials at the 
collocation points $T_{kn}$ and their first and second derivatives $T'_{kn}$ 
and $T''_{kn}$ form dense real matrices of dimensions $(N_r\times N_r)$ that 
are precalculated and stored in the initialisation procedure of the code.
In \texttt{pizza}, the discretised equations 
(\ref{eq:psiomcoll}-\ref{eq:tempcoll}) supplemented by the boundary conditions 
(\ref{eq:bcs_coll_dirichlet}) or (\ref{eq:bcs_coll_neumann}) are solved using 
\texttt{LAPACK}\footnote{\url{http://www.netlib.org/lapack/}}. The LU 
decomposition is handled by the routine \texttt{dgetrf} or its 
complex-arithmetic counterpart \texttt{zgetrf} and require $\mathcal{O}(N_r^3)$ 
operations per Fourier mode $m$.  This needs to be done at the initialisation 
stage of the code or at each iteration where a change in the time-step size 
occurs (see \S~\ref{sec:tschemes}). During each time step, the routines 
\texttt{dgetrs} (or 
\texttt{zgetrs}) are employed for the matrix solve and correspond to 
$\mathcal{O}(N_r^2)$ operations per Fourier mode $m$. The amount of memory 
required to store the 
dense complex-type matrix that enters the left-hand-side of 
Eq.~(\ref{eq:psiomcoll}) grows as $64\,N_r^2$ for one single azimuthal 
wavenumber $m$ for a double-precision calculation. This corresponds to 
1~Gigabyte of memory per Fourier mode for $N_r=4096$ and hence makes the 
collocation approach extremely costly when  $N_r \gtrsim 10^3$.

\subsection{Spectral equations using a Chebyshev integration method}

To circumvent the limitations inherent in the collocation approach, several 
efficient Chebyshev spectral methods have been developed  
\citep[e.g.][]{Coutsias96,Julien09,Olver13}. They all involve the solve of 
sparse matrices that are almost banded and can be inverted in 
$\mathcal{O}(p\,N_r)$ operations, $p$ being the number of bands of the matrices.
One approach, first introduced by \cite{Clenshaw57}, consists of integrating 
$q$ times a set of $q$th-order ordinary differential equations (ODEs) in 
Chebyshev space \citep[see also][]{FoxParker68,Phillips90,Greengard91}. First 
limited to ODEs with constant coefficients, this method has been further 
extended by \cite{Coutsias96} to ODEs with rational function coefficients. The 
comparison of several Chebyshev methods for fourth-order ODEs carried out by 
\cite{Muite10} showed the advantages of such a Chebyshev integration method 
both in terms of matrix condition number and computational cost in the limit of 
large $N_r$. This technique has been successfully applied to the problem of 
rotating convection in both Cartesian \citep{Stellmach08} and spherical 
geometry \citep{Marti16}.

\subsubsection{Semi-discrete formulation}

The Chebyshev integration methodology relies on the following
indefinite integral identity \citep[e.g.][Eq.~2.4.23]{CHQZ}
\begin{equation}
 \int T_n(x) \mathrm{d}x = 
\dfrac{1}{2}\left[\dfrac{T_{n+1}(x)}{n+1}-\dfrac{T_{n-1}(x)}{n-1}\right]~
\text{for}~n > 1,
\label{eq:int1}
\end{equation}
which in its discrete form corresponds to the following sparse operator
\[
 \widehat{\mathcal{I}}_{kn} = 
-\dfrac{1}{2k}\delta_{k+1,n}+\dfrac{1}{2k}\delta_{k-1,n}
 ~\text{for}~k > 1,
\]
where $\delta$ corresponds to the Kronecker symbol. Identities for multiple 
integration can then be easily derived by recursive applications of 
Eq.~(\ref{eq:int1}).

Because of the singularity of $\beta$, we first need to regularise the set of 
equation (\ref{eq:psi}-\ref{eq:temp}) to make it suitable for a Chebyshev 
integration method. We hence adopt the following different definition for the 
streamfunction $\varPsi$
\[
 u_s = \dfrac{1}{s}\dfrac{\partial  [\zeta(s) \varPsi]}{\partial \phi}; \quad 
u_\phi = \overline{u_\phi}
-\dfrac{\partial [\zeta(s) \varPsi]}{\partial s}-\beta \zeta(s)\varPsi\,.
\]
Using $\zeta(s) =h^2= s_o^2-s^2$ then yields
\begin{equation}
 u_s = \dfrac{h^2}{s} \dfrac{\partial \varPsi}{\partial \phi};\quad
 u_\phi = \overline{u_\phi}-h^2\dfrac{\partial \varPsi}{\partial 
s}+3s\,\varPsi\,.
\end{equation}
From these definitions, one derives the following expression for the 
axial vorticity 
$\omega$
\begin{equation}
 \omega =\dfrac{1}{s}\dfrac{\partial (s\overline{u_\phi})}{\partial s} 
-\mathcal{L}_I \varPsi\,,
\end{equation}
where the operator $\mathcal{L}_I$ is given by
\[
\mathcal{L}_I \varPsi=\Delta\left(h^2
\varPsi\right)-\dfrac{1}{s}\dfrac{\partial}{\partial 
s}\left(s^2 \varPsi \right)\,.
\]
The expansion of $\varPsi$ and $\vartheta$ in Fourier modes 
yields the following equation for the time evolution of $\varPsi$ for the 
non-axisymmetric Fourier modes
\[
 \left[\left(\dfrac{\partial}{\partial 
t}-\Delta\right)\mathcal{L}_I-\dfrac{2}{E}\mathrm{i}m\right]\varPsi_m = 
\dfrac{Ra}{Pr}\dfrac{\mathrm i m}{s_o}\vartheta_m+\mathcal{N}_{\omega 
m}-\mathcal{F}_\epsilon(E,\varPsi_m)\,.
\]
In the above equation, the classical Ekman pumping term 
(Eq.~\ref{eq:pumping_full}) has been replaced by the 
approximated form $\mathcal{F}_\epsilon$ 
defined by
\begin{equation}
 \mathcal{F}_\epsilon 
=\Upsilon_\epsilon \left[\mathcal{L}
_I+\dfrac{s}{2}\dfrac { \partial} {\partial 
s}-\left(\dfrac{3s^2}{2h_\epsilon^2}+m^2+\dfrac{5\mathrm{i} ms_o}{ 
2h_\epsilon}\right)
\right]\varPsi_m
\label{eq:approx_pump}
\end{equation}
where $h_\epsilon= [(s_o+\epsilon)^2-s^2]^{1/2}$ corresponds to half the height 
of a geostrophic cylinder that would intersect a sphere with a slightly larger 
radius $s_o+\epsilon$, with $\epsilon \ll 1$. $\Upsilon_\epsilon$ is 
defined accordingly by $\Upsilon_\epsilon= s_o^{1/2}/E^{1/2}/h_\epsilon^{3/2}$ .
This implies that $\mathcal{F}_\epsilon$ corresponds to the exact Ekman 
pumping contribution that would occur in a spherical QG set-up with an outer 
radius $s_o+\epsilon$. In other words, the approximated Ekman pumping  
$\mathcal{F}_\epsilon$ tends to approach the exact contribution
$\mathcal{F}$ in the limit of vanishing $\epsilon$.
This approximation is required when using a Chebyshev integration method to 
avoid the outer boundary singularity of the exact Ekman pumping term and to get 
a good spectral representation of this quantity once transformed 
to Chebyshev space. The error introduced by this approximation will be further 
assessed in \S~\ref{sec:results}.

In addition, the Ekman pumping term requires special care 
since it comprises non-rational function coefficients. In contrast to the 
collocation method where it can be treated implicitly without any additional 
cost, this term shall hence be treated as yet another non-linear 
term since its implicit treatment would yield a dense operator with the 
Chebyshev integration method \citep{Hiegemann97}.

The equation for the time evolution of $\varPsi$ is regularised by a 
multiplication by $s^4$ and then integrated four times to yield
\begin{equation}
\begin{aligned}
 \int\!\!\!\!\int\!\!\!\!\int\!\!\!\!\int 
s^4 \left[\left(\dfrac{\partial}{\partial 
t}-\Delta\right)\mathcal{L}_I-\dfrac{2}{E}\mathrm{i}m\right]\varPsi_m =
\alpha r^3+\beta r^2+\gamma r +\delta \\ 
+\int\!\!\!\!\int\!\!\!\!\int\!\!\!\!\int s^4 \left[
\dfrac{Ra}{Pr}\dfrac{\mathrm i m}{s_o}\vartheta_m
+\mathcal{N}_{\omega 
m}-\mathcal{F}_\epsilon(E,\varPsi_m)\right]\,,
\end{aligned}
\label{eq:psiint}
\end{equation}
where $\alpha, \beta, \gamma$ and $\delta$ are constant of integration that 
will not be required once this equation has been supplemented by boundary 
conditions. At this 
stage, any single term that enters the above equation can be written as the 
product $x^q \partial^p f /\partial x^p$, where $p$ and $q$ are positive 
integers. Following \cite{Marti16}, this equation is 
then integrated by parts until no differential operator remains, such that each 
term has the following form
\[
 \sum_{p=0}^4 \underbrace{\int\cdots\int}_{p\times} \left(\sum_q  x^q f(x) 
\right)\mathrm{d}x^p\,.
\]
After expanding $f(x)$ in Chebyshev polynomials using Eq.~(\ref{eq:cheb}), the 
semi-discrete representation of Eq.~(\ref{eq:psiint}) can be derived by 
multiple application of the recurrence relation (\ref{eq:int1}). This yields

\begin{equation}
\begin{aligned}
\sideset{}{''}\sum_{n=0}^{N_c-1}
\left(\dfrac{\mathrm{d} }{\mathrm{d} t}  
\mathcal{A}^I_{mkn}-\mathcal{B}^I_{mkn}\right)
\widehat{\varPsi}_{mn}(t) = \\
\sideset{}{''}\sum_{n=0}^{N_c-1}
\mathcal{C}^I_{kn} 
\left[\dfrac{Ra}{Pr}\dfrac{\mathrm{i}m}{s_o}\widehat{\vartheta}_{mn}(t)+\widehat
{
\mathcal { N }}_{\omega 
mn}-\widehat{\mathcal{F}}_{\epsilon\,n}(E,\varPsi_m)\right],
\end{aligned}
\label{eq:psi_int}
\end{equation}
for $k>4$. 
$\mathcal{A}^I_{mkn}$, $\mathcal{B}^I_{mkn}$, and 
$\mathcal{C}^I_{kn}$ are the discrete representations of the following operators
\[
\begin{aligned}
\mathcal{A}^I_m = \int\!\!\!\!\int\!\!\!\!\int\!\!\!\!\int s^4 \mathcal{L}_I;\ 
& \mathcal{B}^I_m = \int\!\!\!\!\int\!\!\!\!\int\!\!\!\!\int s^4 \left(\Delta 
\mathcal{L}_I+\dfrac{2}{E}\mathrm{i} m\right); \\
&\mathcal{C}^I =\int\!\!\!\!\int\!\!\!\!\int\!\!\!\!\int s^4
\end{aligned}
\]
The internal matrix elements are determined using the freely available 
\texttt{python} package developed by 
\cite{Marti16}\footnote{It can be downloaded as part of the supplementary 
materials of the study by \cite{Marti16} \href{%
https://agupubs.onlinelibrary.wiley.com/action/downloadSupplement?doi=10.1002\%2F2016GC006438&file=ggge21074-sup-0002-2016GC006438-s02.zip}{here}.} 
that allows the symbolic computation of those 
operators\footnote{\url{https://www.sympy.org/}}. Excluding boundary 
conditions, $\mathcal{A}^I_m$, $\mathcal{B}^I_m$ and $\mathcal{C}^I$ correspond 
to band matrices with $p_u$ super-diagonals and $p_\ell$ sub-diagonals that 
have a bandwidth defined by 
\[
q=p_\ell+p_u+1\,.
\]
The bandwidth of $\mathcal{A}^I_{m}$, 
$\mathcal{B}^I_m$, and $\mathcal{C}^I$ is 17, 13 and 17, respectively.

We proceed the same way to establish the equations for the axisymmetric zonal 
flow component and for the temperature perturbation. Eq.~(\ref{eq:uphi}) and 
Eq.~(\ref{eq:temp}) are multiplied by $s^2$ and integrated twice to yield
\begin{equation}
\begin{aligned}
\sideset{}{''}\sum_{n=0}^{N_c-1}
\left(\dfrac{\mathrm{d} }{\mathrm{d} t}  
\mathcal{D}^I_{kn}-\mathcal{E}^I_{kn}\right)
\widehat{u_\phi}_{0n}(t) = \\
-\sideset{}{''}\sum_{n=0}^{N_c-1}
\mathcal{D}^I_{kn}\left[ \widehat{\mathcal{N}}_{u_{\phi}mn} 
+\widehat{\Upsilon_\epsilon {u_\phi}_0}_n\right],
\end{aligned}
\label{eq:uphi_int}
\end{equation}
for the axisymmetric zonal flow component and
\begin{equation}
\begin{aligned}
\sideset{}{''}\sum_{n=0}^{N_c-1}
\left(\dfrac{\mathrm{d} }{\mathrm{d} t}  
\mathcal{D}^I_{kn}-\dfrac{1}{Pr}\mathcal{F}^I_{kmn}\right)
\widehat{\vartheta}_{mn}(t) = \\
-\sideset{}{''}\sum_{n=0}^{N_c-1}
\mathcal{D}_{kn} 
\left[\mathrm{i}m\widehat{\left(\dfrac{h^2}{s}\dfrac{\mathrm{d}T_c}{\mathrm{d}s}
\varPsi_m\right)}_ {n}+\widehat {\mathcal { N }}_{\vartheta 
mn}\right],
\end{aligned}
\label{eq:temp_int}
\end{equation}
for the temperature. Both equations are only valid for $k>2$.
 $\mathcal{D}^I_{kn}$, $\mathcal{E}^I_{kn}$ and 
$\mathcal{F}^I_{mkn}$ are the discrete representation of the following operators
\[
\mathcal{D}^I = \int\!\!\!\!\int s^2;\ 
  \mathcal{E}^I = s^2-3\int s;\ 
\mathcal{F}^I_m = \int\!\!\!\!\int s^2 \Delta\,.
\]
The bandwidth of $\mathcal{D}^I$, $\mathcal{E}^I$ and 
$\mathcal{F}^I_m$ is 9, 5 and 5, respectively.
 In contrast to the semi-discrete equations obtained with the 
collocation approach, the right-hand-sides of 
Eq.~(\ref{eq:psi_int}-\ref{eq:temp_int})
now involve nonlinear terms that are in Chebyshev space. To avoid 
aliasing errors, the Chebyshev coefficients of nonlinear terms that have 
$n> 2N_r/3$ are hence set to zero \citep{Orszag71}.

\subsubsection{Boundary conditions}

%

At this stage, the system of equation (\ref{eq:psi_int}-\ref{eq:temp_int}) 
needs to be supplemented by boundary conditions. Given the 
definition of $\varPsi$, the rigid mechanical boundary conditions 
that require the cancellation of $u_s$ and $u_\phi$ at both 
boundaries are already ensured by the three following identities:
\begin{equation}
 \varPsi(s=s_i)=\dfrac{\partial \varPsi}{\partial s}(s=s_i)=0, \quad  
\varPsi(s=s_o)=0\,.
\label{eq:bc_varPsi}
\end{equation}
An extra boundary condition on $\varPsi$ is thus required. 
Following \cite{Bardsley18}, we make the ansatz
\[
 \varPsi \sim (s_o^2-s^2)^n  \  \text{when}\ s \rightarrow s_o\,.
\]
This yields the following expression for the viscous term
\[
 \Delta \mathcal{L}_I \varPsi = \dfrac{1}{s^4}(s_o^2-s^2)^{n-3}\left[8 n 
\,s_o^8\,(-2n^3+3n^2+5n-6)\right]\,,
\]
when $s\rightarrow s_o$.
A finite solution requires either $n>3$ or  the cancellation of the 
poynomial on $n$, which has four roots $(-3/2,0,1,2)$. $n=-3/2$ is not 
allowed and $n=0$ is redundant with the cancellation of $\varPsi$ at 
$s=s_o$. Hence the first possible solution is $n=1$ which yields
\[
 \varPsi \sim s^2-s_o^2\  \text{when}\ s \rightarrow s_o\,.
\]
This corresponds to the following additional boundary condition
\begin{equation}
 \dfrac{\partial^3 \varPsi}{\partial s^3} = 0\ \text{for}\ s=s_o\,.
 \label{eq:bc_d3psi}
\end{equation}

When using the Chebyshev integration method, the boundary conditions can be 
either enforced via the tau-Lanczos method or by setting up an adapted Galerkin 
basis function \citep{CHQZ,Boyd01}. In the tau-Lanczos formulation, the top 
rows of the matrices are used to enforce the boundary conditions, which are 
actually identical to the ones used in the collocation method
(Eqs.~\ref{eq:bcs_coll_dirichlet}-\ref{eq:bcs_coll_neumann}). The 
fourth condition on $\varPsi$ given in Eq.~(\ref{eq:bc_d3psi}) corresponds to 
the following last tau line \citep[see][]{Julien09}
\begin{equation}
  \sideset{}{''}\sum_{n=0}^{N_c-1} n^2(n^2-1)(n^2-4)\,\widehat{\varPsi}_{n} = 
0\,.
  \label{eq:bc_d3psi_coll}
\end{equation}
Figure~\ref{fig:mat}b shows the structure of the matrix that enters the 
left-hand-side of Eq.~(\ref{eq:psi_int}) when the boundary conditions are 
enforced using a tau-Lanczos formulation. The two first rows of the matrix 
correspond to the Dirichlet boundary conditions 
(Eqs.~\ref{eq:bcs_coll_dirichlet} and \ref{eq:bc_varPsi}), the third one to 
the above equation and the fourth one to the Neumann boundary condition  
(Eqs.~\ref{eq:bcs_coll_neumann} and \ref{eq:bc_varPsi}). Below those four full 
lines the matrix has a banded structure with 8 sub- and super-diagonals. This 
corresponds to a so-called bordered matrix wich can be inverted in 
$\mathcal{O}(17\,N_r)$ operations as long as the number of full rows is small
compared to the problem size \citep[e.g.][]{Boyd01}. Appendix~\ref{sec:app1} 
gives the details of the matrix inversion procedure as implemented in 
\texttt{pizza}.

We proceed the same way for the boundary conditions on the axisymmetric zonal 
flow and on the temperature. In those cases the Dirichlet boundary conditions
(\ref{eq:bcs_coll_dirichlet}) are imposed as the two first tau lines of the 
matrix, while the banded structure below is given by (\ref{eq:uphi_int}) and 
(\ref{eq:temp_int}), respectively.

Alternatively, the boundary conditions can be imposed by introducing a suitable
Galerkin basis. The underlying idea is to define basis 
functions that satisfy the boundary conditions such that the solutions 
expressed on this set of functions will also directly fulfill the boundary 
conditions. The Galerkin basis of functions $\phi_m$ is usually defined as a 
linear combination of a small number $n_c$ of Chebyshev polynomials
\[
 \phi_n(x) = \sum_{i=0}^{n_c-1} \gamma_i^n T_{n+i}(x)\,.
\]
We first construct the Galerkin basis for the four boundary conditions on 
$\varPsi$ (Eqs.~\ref{eq:bc_varPsi} and \ref{eq:bc_d3psi}). Following 
\cite{Julien09}, the tau conditions (\ref{eq:bcs_coll_dirichlet}, 
~\ref{eq:bcs_coll_neumann}, \ref{eq:bc_d3psi_coll}) are used to establish a 
related Galerkin set. Appendix~\ref{sec:app2} gives the details of the 
calculation of the $\gamma_i^n$ coefficients for $0\leq i \leq 4$.
$\varPsi$ is then decomposed on the Galerkin basis as follows
\[
 \varPsi(s) = \sum_{n=0}^{N_r-5} \widetilde{\varPsi}_n \phi_n(x)\,,
\]
where the tilda notation denotes the Galerkin 
coefficients. The Galerkin coefficients $\widetilde{\varPsi}$ relate to the 
Chebyshev  coefficients $\widehat{\varPsi}$ via
\[
 \widehat{\varPsi} = S_\varPsi\,\widetilde{\varPsi},
\]
where $S_\varPsi$ is the stencil matrix that contains the coefficients 
$\gamma_i$. For 
the Galerkin basis employed for the equation on $\varPsi$, $S_\varPsi$ is a 
band matrix 
with four sub-diagonals. The Galerkin formulation of Eq.~(\ref{eq:psi_int}) 
can be hence written in its matrix form as
\begin{equation}
 P_4 \left(\dfrac{\mathrm{d} \mathcal{A}^I_m }{\mathrm{d} 
t}-\mathcal{B}^I_m\right) 
S_\varPsi 
\,\widetilde{\varPsi}_m = P_4
C^I\left[\dfrac{Ra}{Pr}\dfrac{\mathrm{i}m}{s_o}\widehat{\vartheta}_{m}+\widehat{
\mathcal { N }}_{\omega 
m}-\widehat{\mathcal{F}}_{\epsilon}\right],
\label{eq:psiint_galerkin}
\end{equation}
where $P_4$ is an operator that removes the top four rows of the matrices, which
correspond to the number of boundary conditions \citep{Julien09}.
Figure~\ref{fig:mat}c shows the structure of the matrix that enters the 
left-hand-side of Eq.~(\ref{eq:psiint_galerkin}). Compared to the bordered 
matrix obtained 
when using the tau method, the matrix has now a pure banded structure with an 
increased bandwidth with 8 sub- and 12 super-diagonals. Those matrices 
could be solved using standard band matrix solvers. In \texttt{pizza}, the LU 
decomposition is handled by the \texttt{LAPACK} routine \texttt{dgbtrf} or its 
complex arithmetic counterpart \texttt{zgbtrf} in $\mathcal{O}(q^2\,N_r)$ 
operations per Fourier mode $m$. \texttt{dgbtrs} (or \texttt{zgbtrs})  
routines are then employed for the matrix solve in $\mathcal{O}(q\,N_r)$ 
operations per Fourier mode $m$.

We proceed the same way for the zonal velocity and the 
temperature equations by defining a Galerkin basis that 
ensures Dirichlet boundary conditions at both boundaries.
Several different Galerkin basis sets that 
satisfy this type of boundary conditions have been frequently used in the 
context of modelling rotating convection \citep[e.g.][]{Pino00,Stellmach08}. 
Following \cite{Julien09}, we decide here to adopt the following set
\begin{equation}
 \phi_n(x)=T_{n+2}(x)-T_n(x),\quad\text{for}\quad n<N_r-3\,.
 \label{eq:galerkin_dirichlet}
\end{equation}
In matrix form, the Galerkin formulations of equations 
(\ref{eq:uphi_int}) and (\ref{eq:temp_int}) yield
\begin{equation}
P_2\left(\dfrac{\mathrm{d} \mathcal{D}^I}{\mathrm{d} t}  
-\mathcal{E}^I\right) S_{\text{D}}\,
\widetilde{u_\phi}_{0} = \\
-P_2
\mathcal{D}^I\left[ \widehat{\mathcal{N}}_{u_{\phi}m} 
+\widehat{\Upsilon_\epsilon {u_\phi}_0}\right],
\label{eq:uphi_galerkin}
\end{equation}
for the axisymmetric zonal flow component and
\begin{equation}
P_2\left(\dfrac{\mathrm{d} D^I }{\mathrm{d} t}  
-\dfrac{\mathcal{F}^I_m}{Pr}\right)S_{\text{D}}\,
\widetilde{\vartheta}_m  = -P_2
\mathcal{D}^I 
\left[\mathrm{i}m\widehat{\left(\dfrac{h^2}{s}\dfrac{\mathrm{d}T_c}{\mathrm{d}s}
\varPsi_m\right)}+\widehat {\mathcal { N }}_{\vartheta m}\right],
\label{eq:temp_galerkin}
\end{equation}
for the temperature, where $S_\text{D}$ is the stencil matrix 
(\ref{eq:galerkin_dirichlet}) and 
$P_2$ is an operator that removes the top two rows.

We note that different type of boundary conditions, such as stress-free and/or 
fixed flux thermal boundary conditions, would necessitate the derivation of 
dedicated Galerkin bases following a procedure similar to the one 
discussed in the appendix~\ref{sec:app2}.

Previous analysis by \cite{Julien09}  showed that the Galerkin approach
usually yield matrices with a better condition number than the bordered 
matrices obtained when using the tau-Lanczos method. This is particularly 
critical when 2-D or 3-D Chebyshev domains are considered but remains 
acceptable for 1-D problem as considered here \citep[see 
Table~1 in][]{Julien09}. The Galerkin approach should hence 
be privileged as long as homogeneous boundary conditions are enforced, while
inhomogeneous boundary conditions for which a Galerkin description becomes 
cumbersome are easier to handle with a tau-Lanczos formulation.

\section{Temporal discretisation}

\label{sec:tschemes}

\begin{table*}
 \caption{Time schemes implemented in \texttt{pizza}. The 
fifth ($\mathcal{I}$) and the sixth columns ($\mathcal{E}$) correspond 
to the number of implicit and explicit terms computed for one time step, 
respectively. The seventh column (Storage) is the number of state vectors that 
need to be stored to time-advance one physical quantity. The eighth column 
(Cost) corresponds to the elapsed wall time for one iteration normalised by 
the cost for one iteration of CNAB2. The last column contains the maximum CFL
$\alpha$ obtained for a case with $E=10^{-7}$, $Ra=2\times 10^{11}$, $Pr=1$ and 
$\epsilon=10^{-3}$ which has been computed using the Chebyshev integration and 
Galerkin methods with $(N_r,N_c,N_m)=(1025,682,1280)$. The asterisks 
corresponds to the models which have been run with an explicit treatment of the 
buoyancy term.}
  \centering
 \begin{tabular}{lllccccccc}
   \toprule
   Name &  Family &Reference & Order & $\mathcal{I}$ &$\mathcal{E}$ & Storage & 
Cost &$\alpha$ \\
 	\midrule
    SBDF4 & Multi-step &\cite{Wang08}, Eq.~(2.15) & 4 & 1&1 & 8 & 1.01 &0.19\\
    SBDF3 & Multi-step &\cite{Peyret02}, Eq.~(4.83) & 3 & 1&1 & 6 
& 0.97 &0.23\\
    SBDF2 & Multi-step &\cite{Peyret02}, Eq.~(4.82) & 2 & 1&1 & 4 & 0.96 &0.21\\
    CNAB2 & Multi-step &\cite{Glatzmaier84}, Eq.~(5b) & 2 & 1&1 & 4 & 1 &0.25\\
	\midrule
    BPR353 & SDIRK & \cite{Boscarino13}, \S~8.3 & 3 & 5&3 & 9 & 3.24 &$0.78^*$\\
    ARS443 & SDIRK & \cite{Ascher97}, \S~2.8 & 3 & 4&3 & 9 & 3.66 &$0.71^*$\\
    ARS222 & SDIRK & \cite{Ascher97}, \S~2.6 & 2 & 2&2 & 5 & 1.81 &$0.45^*$\\
    LZ232  & SDIRK & \cite{Liu06}, \S~6 & 2 & 2&2 & 6 & 1.86 &$0.42^*$\\
   \bottomrule
 \end{tabular}
 \label{tab:timeschemes}
\end{table*}

The equations discretised in space can be written as a general 
ordinary differential equation in time where the right-hand-side is split in two 
contributions
\begin{equation}
 \dfrac{\mathrm{d} y}{\mathrm{d} t} = \mathcal{E}(y,t) + \mathcal{I}(y,t), 
\quad 
y(t_0)=y_0,
 \label{eq:EDP}
\end{equation}
where $\mathcal{I}(y,t)$ corresponds to the linear terms, while 
$\mathcal{E}(y,t)$ corresponds to the nonlinear advective 
terms. Temporal stability constraints coming from the linear terms that enter 
Eqs.~(\ref{eq:vort}-\ref{eq:temp}) is usually more stringent that the one 
coming from the nonlinear terms. Except for weakly nonlinear 
calculations, this precludes the usage of purely explicit time schemes such as 
the popular fourth order Runge-Kutta \citep[e.g.][]{Grooms11}. Although they 
offer an enhanced stability, purely implicit schemes are extremely costly since 
they involve the coupling of all Fourier modes due to the implicit treatment of 
the nonlinear terms. The potential gain in time step size is hence cancelled by 
the numerical cost associated with the solve of large matrices.
In the following, we hence only consider \emph{implicit-explicit} 
schemes (hereafter IMEX) to solve Eq.~(\ref{eq:EDP}) and to produce the 
numerical approximation $y_n \simeq y(t_n)$. We first consider the 
general $k$-step IMEX linear multistep scheme
\begin{equation}
 y_{n+1} =  \sum_{j=1}^{k} a_j y_{n+1-j}+\delta t \left(\sum_{j=1}^{k} 
b^\mathcal{E}_j \mathcal{E}_{n+1-j}+\sum_{j=0}^{k} b^\mathcal{I}_j 
\mathcal{I}_{n+1-j}\right),
 \label{eq:multistep}
\end{equation}
where $\mathcal{E}_{n+1-j}=\mathcal{E}(y_{n+1-j},t_{n+1-j})$ and 
$\mathcal{I}_{n+1-j}=\mathcal{I}(y_{n+1-j},t_{n+1-j})$. The vectors 
$\vec{a}$, $\vec{b}^\mathcal{E}$ and $\vec{b}^\mathcal{I}$ correspond to the 
weighting 
factors of the IMEX multistep scheme. For instance, the commonly-used 
second-order scheme assembled from the combination of a Crank-Nicolson for the 
implicit terms and a second-order Adams-Bashforth for the explicit terms 
(hereafter CNAB2) corresponds to the following vectors $\vec{a} = (1,0)$, 
$\vec{b}^\mathcal{I} = (1/2, 1/2)$ and $\vec{b}^\mathcal{E}=(3/2,-1/2)$ 
for a constant
$\delta t$. In practice, Eq.~(\ref{eq:multistep}) is rearranged as follows
\begin{equation}
\begin{aligned}
  (I-b_0^\mathcal{I}\delta t\,\mathcal{I})\,y_{n+1} = & \sum_{j=1}^{k} a_j 
y_{n+1-j}\\&+\delta t 
\sum_{j=1}^{k} \left(b_j^\mathcal{E} \mathcal{E}_{n+1-j}+b_j^\mathcal{I}
\mathcal{I}_{n+1-j}\right)\,,
\end{aligned}
\label{eq:multistep1}
\end{equation}
where $I$ is the identity matrix. In addition to CNAB2, \texttt{pizza} supports 
several semi-implicit backward differentiation schemes of second, third and 
fourth order that are known to have good stability properties \citep[heareafter 
SBDF2, SBDF3 and SBDF4, see][]{Ascher95,Garcia10}. The interested reader is 
referred to the work by 
\cite{Wang08} for the derivation of the vectors $\vec{a}$, 
$\vec{b}^\mathcal{I}$ and 
$\vec{b}^\mathcal{E}$  when the time step size is variable.  
Table~\ref{tab:timeschemes} summarises the main 
properties of the multistep schemes implemented in \texttt{pizza}.

Multistep schemes suffer from several possible limitations: 
(\textit{i}) when the order is larger than two, they are not self-starting and 
hence require to be initiated with another lower-order starting scheme;
(\textit{ii}) limitations of the time step size to maintain stability is more 
severe for higher-order schemes \citep[e.g.][]{Ascher95,Carpenter05}. In 
contrast, the multi-stage Runge-Kutta schemes are self-starting and frequently 
show a stability region that grows with the order of the scheme.
To examine their efficiency in the context of spherical QG convection, we have 
also implemented in 
\texttt{pizza} several Additive Runge Kutta schemes. For this type of IMEX, 
we restrict ourself to the so-called \emph{Diagonally Implicit Runge Kutta} 
schemes (hereafter DIRK) for which each sub-stage can be solved sequentially.
For such schemes, the equation (\ref{eq:EDP}) is time-advanced from $t_n$ to 
$t_{n+1}$ by solving $\nu$ sub-stages
\begin{equation}
 \left( I - a_{ii}^{\mathcal{I}} \delta t\,\mathcal{I} \right) y_i = 
y_{n}+\delta t 
\sum_{j=1}^{i-1}  \left(a_{i,j}^{\mathcal{E}} \mathcal{E}_j + 
a_{i,j}^{\mathcal{I}}\mathcal{I}_j
\right),\ 1\leq i\leq \nu,
\label{eq:dirks}
\end{equation}
where $y_i$ is the intermediate solution at the stage $i$. Finally the 
evaluation of
\[
 y_{n+1} = y_{n}+\delta t\sum_{j=1}^{\nu}\left (b_j^\mathcal{E}
\mathcal{E}_j+b_j^\mathcal{I}\mathcal{I}_j\right).
\]
allows the determination of $y_{n+1}$. A DIRK scheme with $\nu$ stages can be
represented in terms of the following so-called Butcher tables

\[
\renewcommand\arraystretch{1.2}
\begin{array}{c|c}
 \vec{c}^\mathcal{I} & \mathbf{A}^\mathcal{I} \\
   \hline
 & \vec{b}^\mathcal{I}
\end{array}
=
\begin{array}
{c|ccccc}
c_1^\mathcal{I} & a_{11}^\mathcal{I}\\
c_2^\mathcal{I} & a_{21}^\mathcal{I} & a_{22}^\mathcal{I} \\
\vdots & \vdots & \vdots& \ddots\\
c_\nu^\mathcal{I} & a_{\nu1}^\mathcal{I} & a_{\nu2}^\mathcal{I} &\cdots& 
a_{\nu\nu}^\mathcal{I}\\
\hline
& b_{1}^\mathcal{I} & b_{2}^\mathcal{I} & \cdots & b_\nu^\mathcal{I} 
\end{array}\,,
\]
for the implicit terms, and
\[
\begin{array}{c|c}
 \vec{c}^\mathcal{E} & \mathbf{A}^\mathcal{E} \\
   \hline
 &  \vec{b}^\mathcal{E}
\end{array}
=
\begin{array}
{c|ccccc}
0 & 0\\
c_2^\mathcal{E} & a_{21}^\mathcal{E} & 0 \\
\vdots & \vdots & \vdots& \ddots\\
c_{\nu}^\mathcal{E} & a_{\nu1}^\mathcal{E} & a_{\nu2}^\mathcal{E} &\cdots& 0\\
\hline
& b_{1}^\mathcal{E} & b_{2}^\mathcal{E} & \cdots & b_\nu^\mathcal{E}
\end{array}\,,
\]
for the explicit terms, where zero values above the diagonal have been 
omitted. In the following, we only consider the \emph{stiffly accurate} DIRK 
schemes for which the outcome of the last stage gives the end-result, without 
needing any assembly stage \citep{Ascher97}. This corresponds to 
$b_j^\mathcal{I}=a_{\nu j}^\mathcal{I}$ 
and $b_j^\mathcal{E}=a_{\nu j}^\mathcal{E}$ for $1<j<\nu$. In addition, to 
minimise 
the memory storage which is particularly critical in the Chebyshev collocation 
approach, only the DIRK schemes that 
involve one single matrix storage in the implicit solve are retained, i.e. 
$a_{ii}^\mathcal{I}$ is independent of $i$. The latter restriction corresponds 
to the so-called SDIRK (Singly Diagonally Implicit Runge–Kutta) schemes. In the 
following we discuss the convergence and the stability properties of two 
second order  --ARS222 from \cite{Ascher97} and LZ232 from \cite{Liu06}--; and 
two third order SDIRK schemes --ARS443 from \cite{Ascher97} and BPR353 from 
\cite{Boscarino13}--.

\begin{figure*}
 \centering
 \includegraphics[width=16cm]{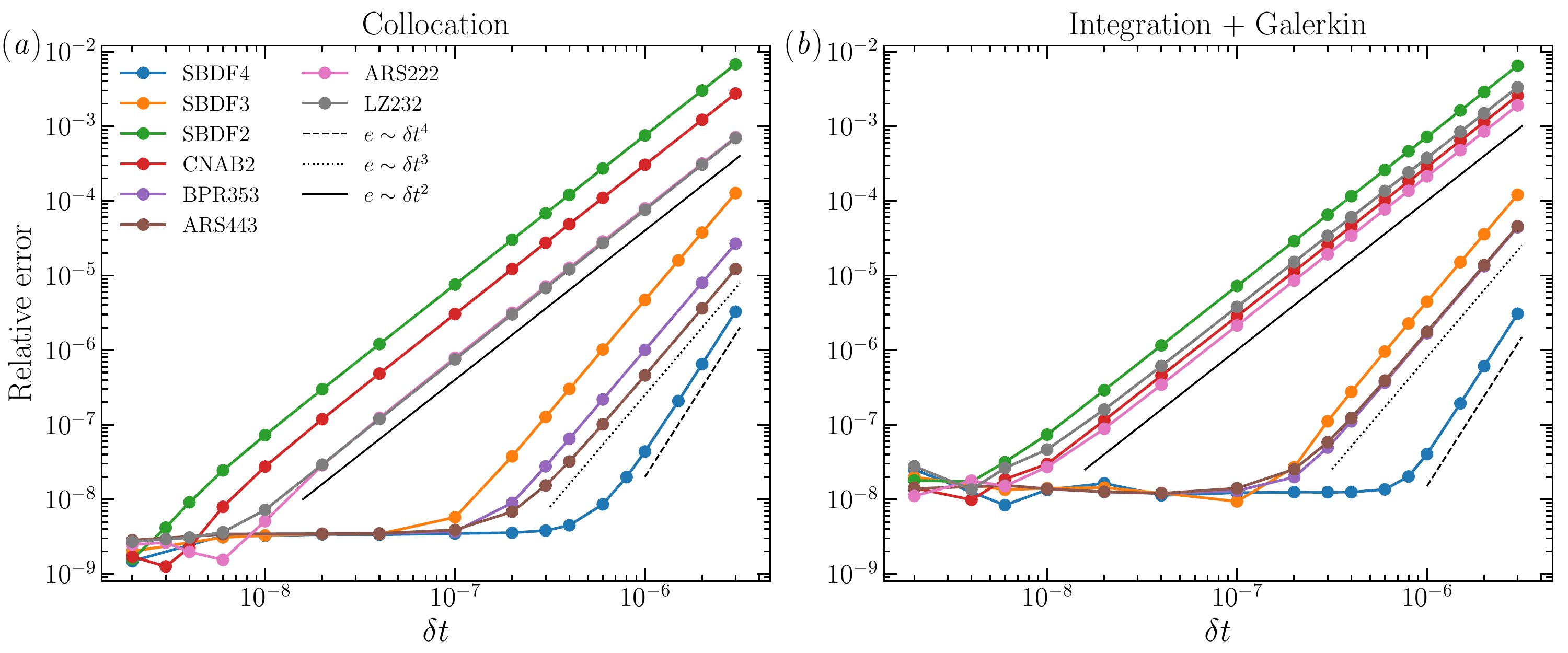}
 \caption{Relative error as a function of the time step size 
$\delta t$ for 
several multistep and SDIRK time schemes when using Chebyshev collocation (left 
panel) and Chebyshev integration method with boundary conditions 
enforced by a Galerkin approach (right panel). For comparison, the 
expected convergence orders have been denoted by black lines. Those convergence 
analyses have been carried out at the saturated stage of a numerical setup with 
$E=3\times 10^{-6}$, $Ra=10^7$ and $Pr=0.025$. Spatial resolution corresponds 
to ($N_r,N_c,N_m=193,193,128$) for the collocation method and 
($N_r,N_c,N_m=193,128,128$) for the integration method. $\epsilon=3\times 
10^{-3}$ has been assumed for the Chebyshev integration method.}
 \label{fig:error_DeltaT}
\end{figure*}

The nonlinear advection terms that enter Eqs.~(\ref{eq:psi}-\ref{eq:uphi}) are 
treated explicitly, while the dissipation terms and the vortex streching term 
in Eq.~(\ref{eq:vort}) are treated implicitly. As long as the fluid domain 
is entirely convecting, the buoyancy term that enters the vorticity equation 
(\ref{eq:vort}) can either be treated explicitly or implicitly without a notable 
change of the stability properties of the IMEX \citep[e.g.][]{Stellmach08}. 
We can expect more significant differences when some regions of 
the fluid are stably stratified. An implicit treatment of the buoyancy term 
only implies that the temperature equation 
(\ref{eq:temp}) shall be first time-advanced to produce
$\vartheta(t_{n+1})$ before time-advancing the vorticity and streamfunction 
\citep[e.g.][]{Glatzmaier84}.
The treatment of the Ekman pumping terms depends on the spatial discretisation 
strategy: while this can be treated implicitly without additional cost in the 
collocation method, this term has to be treated explicitly when using 
the Chebyshev integration method.

For an illustrative purpose, we give here the time-stepping equation for 
$\widehat{\varPsi}_m$ when the Chebyshev integration method 
(Eq.~\ref{eq:psi_int})  
is used  in conjunction with an SDIRK time scheme 
(Eq.~\ref{eq:dirks})
\[
  \begin{aligned}
 \left(\mathcal{A}^I_m -a_{ii}^\mathcal{I} \delta 
t\,\mathcal{B}^I_m\right)\widehat{\varPsi}_m(t_{i}) =
 \mathcal{A}^I_m \widehat{\varPsi}_m(t_n)+\delta t \sum_{j=1}^{i-1} 
a_{i,j}^\mathcal{I}\, 
\mathcal{B}^I_m \widehat{\varPsi}_m(t_{j})\\
  + \delta t \sum_{j=1}^{i-1} a_{i,j}^\mathcal{E}\, 
\mathcal{C}^I\left[\dfrac{Ra}{Pr}\dfrac{\mathrm{i}m}{s_o}\widehat{\vartheta}_{m}
(t_j)+\widehat{\mathcal{N}}_{\omega 
m}(t_j)-\widehat{\mathcal{F}}_{\epsilon}(t_j)\right],
\end{aligned}
\]
where the buoyancy term has been treated explicitly and $1\leq i \leq \nu$. 
This equation needs to be solved $\nu$ times per time step and the outcome of 
the final 
stage produces the time-advanced quantity $\widehat\varPsi_m(t_{n+1})$ for the 
azimuthal wavenumber $m$. A summary of the main properties of the SDIRK schemes 
implemented in \texttt{pizza} is also given in Table~\ref{tab:timeschemes}.

Both families of time integrators (\ref{eq:multistep1}) and (\ref{eq:dirks}) 
have a very similar structure and can hence be implemented using a shared 
framework, provided the programming language supports object-oriented 
implementation \citep{Vos11}. In \texttt{pizza} we rely on the object-oriented 
features provided by the Fortran~2003 norm to implement an abstract framework 
that allows easy switching between different schemes while minimising the 
number of code lines. 

The different time steppers have been validated by running convergence tests.
To do so, we consider a physical test problem with
$E=3\times 10^{-6}$, $Ra=10^7$, $Pr=0.025$ and initiate the numerical 
experiment with a random temperature perturbation. We then run the numerical 
model using an SBDF4 time stepper until a statistically steady-state 
has been reached. This final state serves as the starting conditions of
a suite of numerical simulations that use different fixed time step size 
$\delta t$ 
between $10^{-9}$ and $3\times 10^{-6}$ over a fixed physical timespan 
$t=1.2\times 10^{-3}$. Following \cite{Grooms11},
the error associated with the time stepper is defined as the sum of the 
relative errors on $\vartheta$, $u_s$ and $u_\phi$, where the relative error for 
one physical quantity $f$ is expressed by
\[
 e_{\text{rel}}(f) = \left[\dfrac{\left\langle (f-f_\text{ref})^2 \right\rangle
}{\left\langle f_\text{ref}^2 \right \rangle}\right]^{1/2}\,.
\]
In the above equation, the angular brackets correspond to an integration over 
the annulus
\[
 \langle f \rangle = \int_{0}^{2\pi}\int_{s_i}^{s_o} 
f(s,\phi) \,s\,\mathrm{d}s\,\mathrm{d}\phi\,.
\]
The fourth-order SBDF4 time stepper with the smallest time step size $\delta 
t=10^{-9}$ has been used to define the reference solution $f_{\text{ref}}$.
Figure~\ref{fig:error_DeltaT} shows the error as a function 
of $\delta t$ for the time schemes given in Table~\ref{tab:timeschemes} for 
both the collocation method (left panel) and the Chebyshev integration 
method with a Galerkin approach to enforce the boundary conditions (right 
panel).
All schemes converge with their expected theoretical order until a plateau 
is reached around $3\times 10^{-9}$ for the Chebyshev collocation and $10^{-8}$ 
for the Chebyshev integration method. This can be attributed to the propagation 
of rounding errors that occur in the spectral transforms and in the calculation 
of the radial derivatives \citep{Sanchez04}. In other words, at this level of 
$\delta t$ the error becomes dominated by the spatial 
discretisation errors. For a given order, SDIRK schemes 
are found to be more accurate than their multistep counterparts for the 
majority of the cases.

\begin{figure*}
 \centering
 \includegraphics[width=16cm]{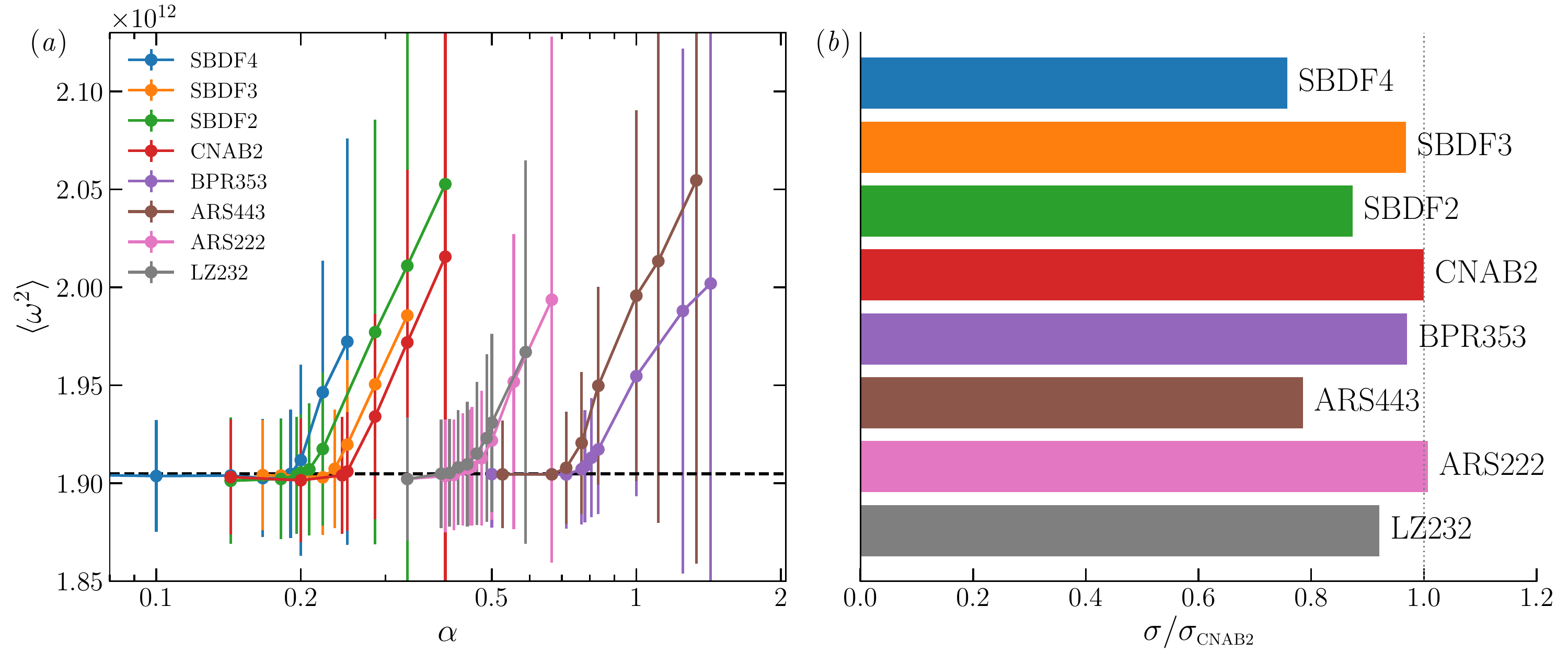}
 \caption{(\textit{a}) Time-averaged total enstrophy $\langle \omega^2 \rangle$ 
as a function of the CFL coefficient $\alpha$ for the time schemes given in 
Table~\ref{tab:timeschemes}. The error bars correspond to one standard deviation 
and the horizontal dashed line corresponds to the time-averaged enstrophy of a 
numerical model computed with the SBDF4 time scheme with $\alpha=0.05$. 
(\textit{b}) Efficiency $\sigma$ of the time 
schemes normalised by the efficiency of CNAB2. The vertical dotted line 
corresponds to a value of one. This efficiency analysis has been carried out 
 at the saturated stage of a numerical model  with $E=10^{-7}$, $Pr=1$, 
$Ra=2\times 10^{11}$. This model was computed using the Chebyshev integration 
and a Galerkin method with a spatial resolution of 
($N_r,N_c,N_m=1025,682,1280$) 
and $\epsilon =10^{-3}$.}
 \label{fig:pizza_crash}
\end{figure*}

This time scheme validation has been carried out with fixed time step sizes 
on a physical test case that is close to the onset of convection. To 
examine the efficiency of the different time schemes to model quasi-geostrophic 
turbulent convection, we also perform a stability analysis on a more turbulent 
setup. Indeed a precision of a fraction of a percent is usually 
sufficient
when considering parameter studies of turbulent rotating convection 
\citep[e.g.][]	{Gastine16}. Hence, the determination of the largest time step 
size 
$\delta t$ is of practical interest to assess the efficiency of a given time 
scheme. To do so, we consider a problem with $E=10^{-7}$, $Ra=2\times 
10^{11}$ and $Pr=1$, which is approximately 60 times supercritical. We 
first time-advance the solution until the nonlinear saturation has been reached 
using a CNAB2 time scheme. We then use the final state of this 
computation as the starting conditions of several numerical simulations that use
different time schemes. Those simulations are computed
over $3\times 10^{-4}$ viscous time, which roughly corresponds to two 
turnover times. Since the advection terms 
are treated explicitly, the maximum eligible time step size must satisfy the 
following Courant criterion
\begin{equation}
 \delta t \leq \alpha \min\left[ \left(\max_{s,\phi} 
\dfrac{|u_s|}{\delta 
s}\right)^{-1}, \left(\max_{s,\phi}\dfrac{|u_\phi|}{s\,\delta\phi}\right)^{-1} 
\right],
 \label{eq:cfl}
\end{equation}
where $\delta s$ correspond to the local spacing of the Gauss-Lobatto 
grid and $\delta \phi= 2\pi/3 N_m$ to the constant spacing in the azimuthal 
direction. 
In the above equation, $\alpha$ corresponds to the
Courant-Friedrichs-Lewy number (hereafter CFL). To determine the CFL number of 
each time scheme, we compute series of simulations with different values of 
$\alpha$  and let the code runs with the maximum allowed $\delta t$ that 
fulfills Eq.~(\ref{eq:cfl}). This implies that $\delta t$ will change at each 
iteration and hence that the matrices will be rebuilt at each time step. 
Since LU factorisation is very demanding when using Chebyshev collocation 
($\mathcal{O}(N_r^3)$ operations), we restrict the stability analysis to the 
sparse Chebyshev integration method with a Galerkin approach to enforce the 
boundary conditions. We use the time evolution of the total enstrophy $\langle 
\omega^2 \rangle$ as a diagnostic to estimate the maximum CFL number $\alpha$.
Because of the clustering of the Gauss-Lobatto grid points, the time step size 
limitation usually occurs in the vicinity of the boundaries. Since  $\langle 
\omega^2 \rangle$ reaches its maximum value in the viscous boundary layers,  
any violation of Eq.~(\ref{eq:cfl}) yields spurious spikes 
in the time evolution of the total enstrophy, well before the code actually 
crashes. For comparison, we define a reference solution that has been run 
with an SBDF4 time scheme with the smallest value of $\alpha = 0.05$.

Figure~\ref{fig:pizza_crash}a shows the time-averaged and the standard 
deviation of $\langle\omega^2\rangle$ as a function of $\alpha$ for the time 
schemes given in Table~\ref{tab:timeschemes}.
The curves are comprised of two parts: one horizontal part where the 
time-averaged total enstrophy remains in close agreement with the reference 
case and the other featuring a rapid increase of both the time-averaged and the 
standard deviation of $\langle \omega^2 \rangle$. We hence define the largest
acceptable  $\alpha$ for a given time scheme as the value above which the 
time-averaged total enstrophy becomes more than 0.3\% larger than the reference 
value. The rightmost column of Table~\ref{tab:timeschemes} documents the 
obtained values. All multi-step schemes exhibit comparable CFL numbers with only 
a weak dependence on the theoretical order of the scheme. This is in agreement 
with the study by \cite{Carpenter05} who report comparable time step limitations 
for several SBDF schemes when the problem becomes numerically stiff. In 
contrast, the SDIRK schemes allow significantly larger CFL numbers with 
third-order schemes being more stable than the second-order ones. We quantify 
the efficiency of a time scheme by the ratio
\begin{equation}
 \sigma = \dfrac{\alpha}{\text{cost}}\,,
 \label{eq:efficiency}
\end{equation}
where the cost corresponds to the average wall time of one iteration without 
LU factorisation (see the before last column in Table~\ref{tab:timeschemes}). 
Figure~\ref{fig:pizza_crash}b shows a comparison of the relative efficiency of 
the time schemes compared to CNAB2. Although the CFL numbers are larger for the 
SDIRK schemes, they actually have a similar efficiency to multistep schemes due 
to their higher numerical cost. CNAB2 and ARS222 are found to be the most 
efficient second-order schemes, while BPR353 and SBDF3 are the best third-order 
schemes. The CFL numbers derived here are however only indicative since the 
stability of the schemes is expected to depend on the stiffness of the physical 
problem \citep[e.g.][]{Ascher97,Carpenter05}. It is yet unclear whether the 
SDIRK schemes considered here will be able to compete with the multistep methods 
in the limit of turbulent quasi-geostrophic convection. Addressing this question
would necessitate a systematic survey of the limits of stability of the 
time schemes over a broad range of Reynolds and Rossby numbers.

\section{Parallelisation strategy}

\label{sec:mpi}

\begin{figure}
 \centering
 \includegraphics[width=8.4cm]{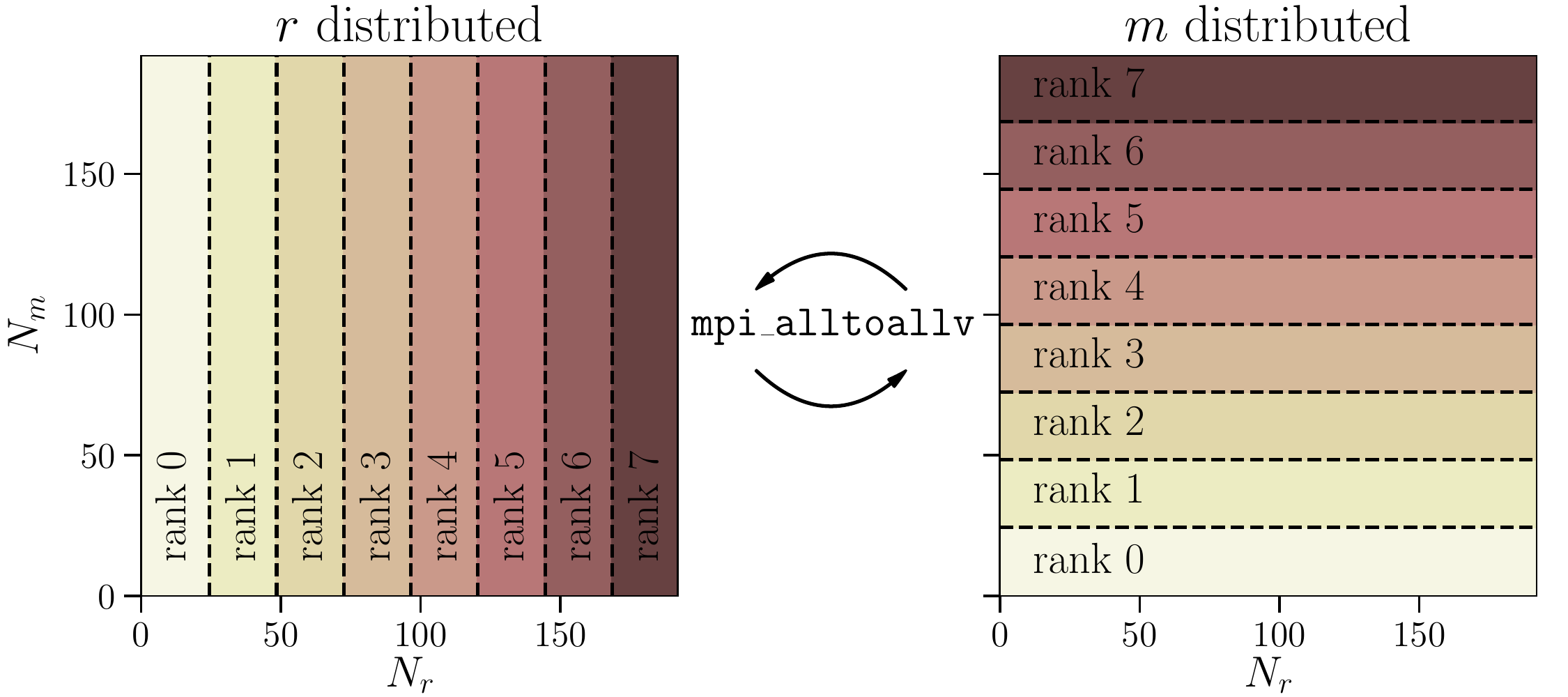}
 \caption{Domain decompositions used in \texttt{pizza}. The left panel 
corresponds to the \texttt{MPI} configuration where the radial levels are 
distributed among ranks and all $m$'s are in processor, while the right panel 
corresponds to the transposed configuration where the azimuthal wavenumbers are 
distributed and all radial level are in processor. The parallel transposition 
between those two  representations is handled by \texttt{mpi\_alltoallv} 
collective communications.}
 \label{fig:mpi_topo}
\end{figure}

The implementation of the algorithm presented before in \texttt{pizza} has 
been designed to run efficiently on massively-parallel architectures. We rely on 
a message-passing communication framework based on the \texttt{MPI} (Message 
Passing Interface) standard. Several approaches have been considered to 
efficiently parallelise spectral transforms between physical and spectral space 
\citep[e.g.][]{Foster97}. Here we decide to resort to a transpose-based 
approach, 
such that all the spectral transforms are applied to data that are local to 
each processor. Whenever needed global transpositions of the data arrays are 
used to ensure that the dimension that needs to be transformed becomes local.

In \texttt{pizza} the data is distributed in two different configurations.
In the first one, the radial level are distributed among \texttt{MPI} ranks 
while all azimuthal wavenumbers are local to each processor. This allows 
the computation of the 1D Fourier transforms (Eq.~\ref{eq:fft}),
the nonlinear terms in the physical space and the backward 
inverse transforms (Eq.~\ref{eq:ifft}). At this stage the data are rearranged 
in 
a second \texttt{MPI} configuration such that the wavenumbers $m$ are 
distributed, while all radial levels are now in processor. Since each processor 
can possibly have a different amount of data to be sent to other processors, 
this parallel transposition is handled by the \texttt{MPI} variant routine 
\texttt{mpi\_alltoallv} that offers dedicated arguments to specify the amount of 
data to be sent and received from each partner. This
configuration is used to time-advance the solution either via Chebyshev 
collocation (Eqs.~\ref{eq:psiomcoll}-\ref{eq:tempcoll}) or  via Chebyshev 
integration method
(Eqs.~\ref{eq:psi_int}-\ref{eq:temp_int}). This implies the solve of linear 
problems and possibly DCTs (Eq.~\ref{eq:cheb}) to transform the data from 
Chebyshev to radial space. Figure~\ref{fig:mpi_topo} summarises the data 
distribution used in \texttt{pizza}.

\begin{figure*}
 \centering
 \includegraphics[width=16cm]{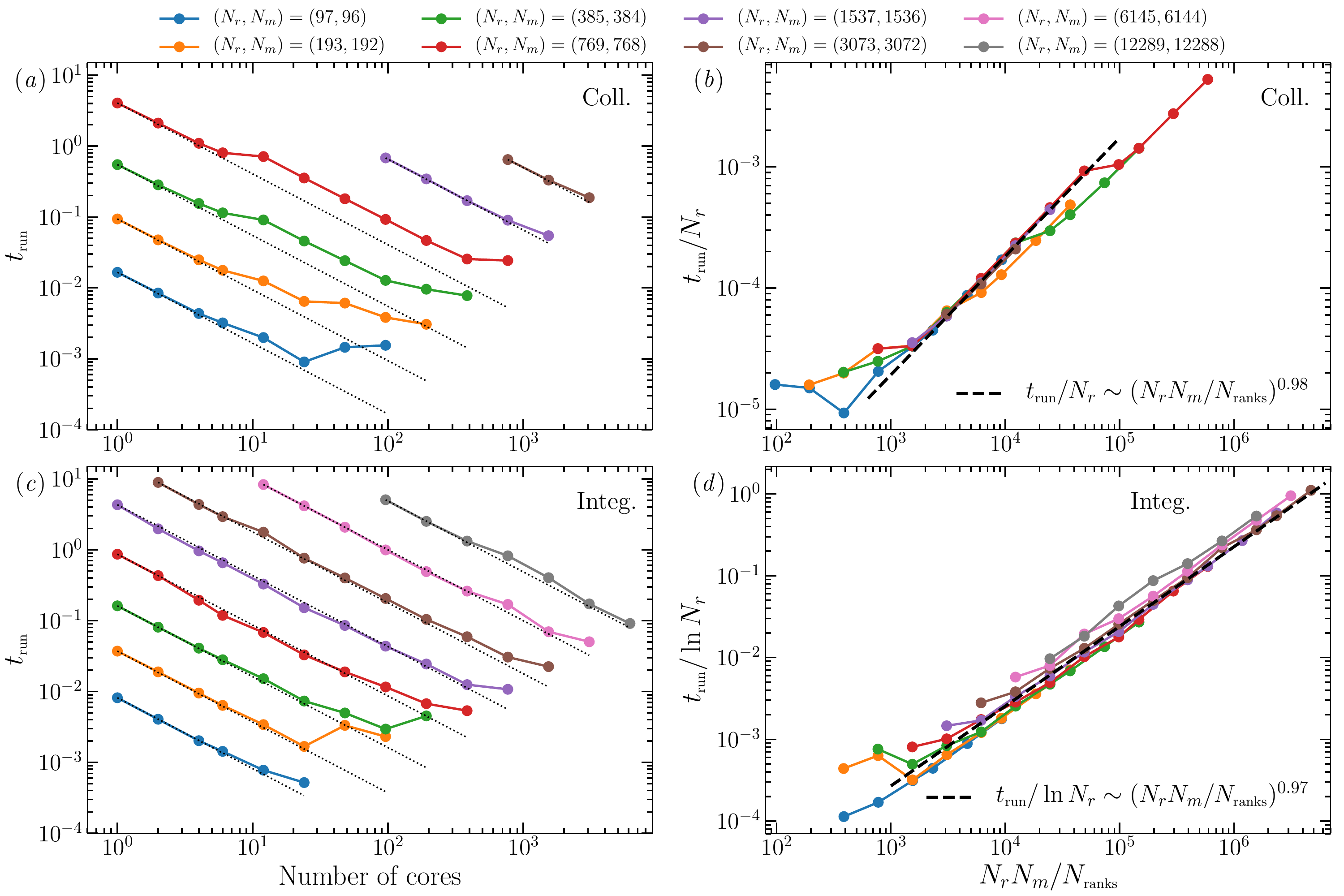}
 \caption{Left panels: wall time per iteration as a function of 
the number of \texttt{MPI} ranks (strong scaling performance) for several 
spatial resolutions. Right panels: wall time per iteration as 
a function of the local data volume per \texttt{MPI} task $N_r 
N_m/N_\text{ranks}$ (weal scaling performance). Panels (\textit{a}) and 
(\textit{b}) corresponds to the models that use the Chebyshev collocation 
method, while panels (\textit{c}) and (\textit{d}) correspond to the models 
where the Chebyshev integration is were used in conjunction with a Galerkin 
approach to enforce the boundary condition. In panels (\textit{a}) and 
(\textit{c}) the dotted black lines correspond to the ideal scalings. The 
linear fits displayed in panels (\textit{b}) and (\textit{d}) have been 
computed from the cases with $(N_r,N_m)=(1537,1536)$. All the 
simulations have been computed using the CNAB2 time scheme.}
 \label{fig:mpi_perf}
\end{figure*}

In the following, we examine the scalability performance of \texttt{pizza}
using the \texttt{occigen} 
cluster\footnote{\url{https://www.cines.fr/calcul/materiels/occigen}}. This 
cluster consists of more than 2000 computational nodes, each node being 
configured with  two Intel 12 cores E5-2690V3 series processor with a clock 
frequency of 2.6~GHz. To build the executable, we make use of the Intel 
compiler version 17.0, Intel \texttt{MPI} version 5.1.3, Intel \texttt{MKL} 
version 17.0 
for the linear solve and the matrix vector products and \texttt{FFTW} version 
3.3.5 for 
Fourier and Chebyshev transforms. We first analyse the strong scaling 
performance of the code by running sequences of numerical simulations with 
several fixed problem size and an increasing number of \texttt{MPI} ranks. The 
left panels in Figure~\ref{fig:mpi_perf} show the wall time per iteration as a 
function of the number of cores for several problem sizes for both Chebyshev 
collocation and Chebyshev integration methods. The resolution $(N_r,N_m)$ range 
from $(97,96)$ to $(12289,12288)$. Because of the dense complex-type matrices 
of size $(2N_r\times 2N_r)$ involved in the time advance of the coupled 
vorticity-streamfunction equation (\ref{eq:psiomcoll}), we cannot use the 
collocation method for the largest problem sizes since it already 
requires more than 1~GB per rank when $N_r=1537$ and $N_m=1536$ with 128 
\texttt{MPI} ranks. For the spatial resolutions that are sufficiently small to 
be computed on one single node, we observe an improved performance when the code 
is running on one single processor (i.e. up to 12 cores) with the Chebyshev 
collocation. This is not observed in the sparse cases and hence might be 
attributed to an internal speed-up of the dense matrix solver of the Intel 
\texttt{MKL} library. Apart from this performance shift, both methods show a 
scalability performance that improves with the problem size. While the 
efficiency of the strong scalings are quickly degraded for $N_{\text{ranks}} > 
N_m/8$ for small problem sizes, \texttt{pizza} shows a very good scalability up 
to $N_{\text{ranks}} = N_m/2$ for the largest problem sizes.
The scalability performance of the collocation method is usually better than the 
Chebyshev integration method for a given problem size. This has to do with the 
larger amount of computational work spent in solving the dense matrices, which 
comparatively reduces the fraction of the wall time that corresponds to
the \texttt{MPI} global transposes. 

In complement to the strong scaling analyses, we also examine weak scaling 
performance tests. This consists of increasing the  number of \texttt{MPI} 
ranks and the problem size accordingly, such that the amount of 
local data per rank stays constant. The spectral transforms implemented in 
\texttt{pizza} require $\mathcal{O}(N_r N_m \ln N_m)$ operations for the FFTs 
(Eq.~\ref{eq:fft}) and $\mathcal{O}(N_m N_r \ln N_r)$ for the DCTs 
(Eq.~\ref{eq:cheb}). The solve of the linear problems involved in the time 
advance of the equations (\ref{eq:psi}-\ref{eq:temp}) grows like 
$\mathcal{O}(N_m N_r^2)$ for the collocation method and only $\mathcal{O}(N_m 
N_r)$ for the Chebyshev integration method. With the 1-D \texttt{MPI} domain 
decomposition discussed above, this implies that an increase of the spatial 
resolution while keeping a fixed amount of local data corresponds to an 
increase of the wall time that should scale with $\mathcal{O}(N_r)$ for the 
collocation method and with $\mathcal{O}(\ln N_r)$  for the Chebyshev 
integration 
method. The right panels of Fig~\ref{fig:mpi_perf} show the wall time per 
iteration normalised by those theoretical predictions as a function of the 
data volume per rank expressed by $N_r N_m/N_\text{ranks}$ for 
both Chebyshev methods. Using the simulations with a spatial resolution of 
$(N_r,N_m)=(1537,1536)$ we compute the following best fits between the 
normalised execution time and the local data volume for each radial 
discretisation scheme

\begin{equation}
  \begin{aligned}
 \dfrac{t_\text{run}^\text{coll.}}{N_r} & = 2.2\times 10^{-8}\left(\dfrac{N_r 
N_m}{N_\text{ranks}}\right)^{0.98}, \\\ 
 \dfrac{t_\text{run}^\text{int.}}{\ln N_r} & = 3.2\times 
10^{-7}\left(\dfrac{N_r 
N_m}{N_\text{ranks}}\right)^{0.97},
\end{aligned}
\label{eq:best_fits}
\end{equation}
where the run time is expressed in seconds. For both methods, the normalised 
wall time per iteration is nearly proportional to the data volume per rank, 
indicating a good agreement with the expected theoretical scalings. We can make 
use of those scalings to estimate the minimum theoretical execution time as a 
function of the problem size. Based on the results of the strong scaling 
analyses, we assume that \texttt{pizza} shows a good parallel efficiency up to 
$N_\text{ranks}=N_m/2$ when the collocation method is used and up to 
$N_\text{ranks}=N_m/4$ when a sparse Chebyshev formulation is employed. 
This yields

\begin{figure}
 \centering
 \includegraphics[width=8.4cm]{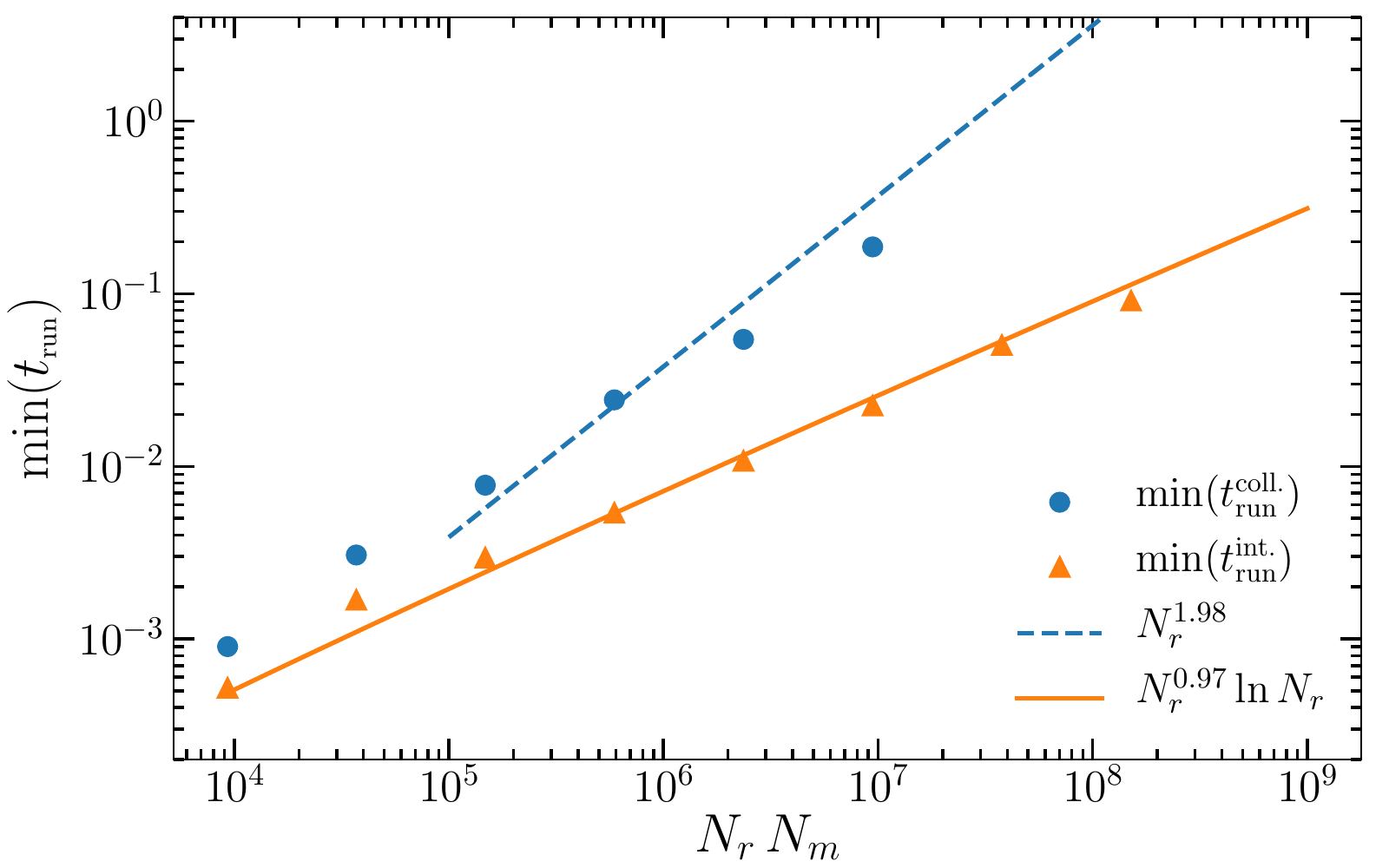}
 \caption{Minimum wall time per iteration as a function of the problem size 
$N_r N_m$. The lines correspond to the linear fits derived from the weak 
scaling tests (see Fig.~\ref{fig:mpi_perf}b and d) for both radial 
discretisation strategies assuming $N_\text{ranks}=N_m/2$ for the collocation 
method and $N_\text{ranks}=N_m/4$ for the Chebyshev integration method 
combined with a Galerkin enforcement of boundary conditions. The symbols 
correspond to the minimum wall times obtained in the strong scaling analyses 
(Fig.~\ref{fig:mpi_perf}a and c).}
 \label{fig:mpi_min_walltime}
\end{figure}

\begin{equation}
\begin{aligned}
 \min(t_\text{run}^\text{coll.}) &= 4.4\times 10^{-8}\, N_r^{1.98}\,, \\
 \min(t_\text{run}^\text{int.}) &= 1.2\times 10^{-6}\,N_r^{0.97} \ln N_r\,.
\end{aligned}
\label{eq:walltimes}
\end{equation}
Figure~\ref{fig:mpi_min_walltime} shows a comparison between the actual 
minimum wall times for different spatial resolutions (see 
Fig.~\ref{fig:mpi_perf}) and the above scalings. A good agreement is found for 
the sparse Chebyshev formulation and for the collocation method with $N_r N_m > 
10^5$. Since the computational time of FFTs and DCTs 
still represents a significant fraction of one time step for small problem 
sizes, this is not surprising that the scaling given in 
Eq.~(\ref{eq:walltimes}) is only approached for sufficiently large problem 
sizes when the collocation method is employed. 

Adopting a Chebyshev integration formulation for the radial scheme provides a 
significant speed up over the collocation approach, with for instance a factor 
10 gain when $N_r N_m \simeq 10^7$. Furthermore, while the collocation method 
becomes intractable for problem sizes with $N_r N_m > 10^7$ because of its 
intrinsic large memory prerequisite, the sparse formulation can be employed 
for spatial resolution larger than $10^4\times 10^4$. Global synchronisation 
and file lock contention can become an issue when reaching this 
range of problem sizes. In \texttt{pizza} this is remedied by collective calls 
to \texttt{MPI-IO} write operations to handle the outputting of checkpoints 
and snapshots.

\section{Code validation and examples}

\label{sec:results}

\subsection{Weakly-nonlinear convection}

In absence of a documented benchmark of spherical QG convection, we test the 
numerical implementation by first looking at the onset of convection. The 
underlying idea being to compare the results coming from a linear eigensolver
with the results from \texttt{pizza}. The comparison of 
the different radial discretisation strategies is of particular interest to 
quantify the error introduced by the approximation of the Ekman pumping term 
involved in the sparse formulation (Eq.~\ref{eq:approx_pump}). To determine the 
onset of spherical QG convection, we linearise the system of equation 
(\ref{eq:psi}-\ref{eq:temp}) and seek for normal modes with 
\[
 f(s,\phi,t) = \Re \left( \sum_{m=0}^\infty f_m(s) e^{\mathrm{i} m 
\phi+\lambda t}\right)\,,
\]
where ${f_m}=(\psi_m,\vartheta_m)^T$ and $\lambda=\tau+\mathrm{i}\omega_d$, 
$\tau$ being the growth rate and $\omega_d$ the angular frequency. Since there 
is no 
coupling between the Fourier modes, we can seek for the solution $f_m$ of one  
individual azimuthal wavenumber. This forms the following generalised
eigenvalue problem
\begin{equation}
 \begin{aligned}
  \lambda \mathcal{L}_\beta \psi_m & = \dfrac{Ra}{Pr}\dfrac{\mathrm{i}m 
}{s_o} \vartheta_m -\dfrac{2}{E}\dfrac{\mathrm{i}m}{s}\beta\psi_m- 
\mathcal{F}(E,\psi_m)+ \Delta (\mathcal{L}_\beta \psi_m)\,, 
\\
  \lambda \vartheta_m & = \Delta \vartheta_m-\dfrac{\mathrm{i} 
m}{s}\dfrac{\mathrm{d}T_c}{\mathrm{d}s}\,\psi_m\,,
 \end{aligned}
 \label{eq:lin}
\end{equation}
that is supplemented by the boundary conditions (\ref{eq:bcs_psi_intro}).
We solve this generalised eigenvalue problem using the \texttt{Linear Solver 
Builder} package (hereafter 
\texttt{LSB}) developed by \cite{Valdettaro07}. The linear operators 
that enter Eq.~(\ref{eq:lin}) are discretised on the Gauss-Lobatto grid using a 
Chebyshev collocation method in real space \citep[e.g.][]{CHQZ}. The entire 
spectrum of complex eigenvalues $\lambda$ is first computed using the QZ 
algorithm \citep{Moler73}. One selected eigenvalue can then be used as 
a guess to accurately determine the closest eigenpair using the iterative 
Arnoldi-Chebyshev algorithm \citep[e.g.][]{Saad92}. As indicated in 
Table~\ref{tab:onset_Gillet}, the linear solver has been 
tested and validated against published values of critical Rayleigh numbers 
for spherical QG convection with or without Ekman pumping \citep{Gillet07}.

\begin{table}
\centering
\caption{Onset of convection for $E=10^{-6}$, $Pr=0.025$ and $r_i/r_o=4/11$ 
from \cite{Gillet07} and obtained with the \texttt{Linear Solver 
Builder} package. Note that the 
critical Rayleigh number $Ra_c$ from \cite{Gillet07} have been normalised to 
match our definition.}
\begin{tabular}{lccc}
 \toprule
& $Ra_c$ & $m$ & $\omega_d$ \\
 \midrule
\multicolumn{4}{c}{Without Ekman pumping} \\
\texttt{LSB} & $1.3851\times 10^7$ & $13$ & $-1.3028\times 
10^4$ \\
\cite{Gillet07} & $1.39\times 10^7$ & $13$ & $-1.300\times 10^4$ \\  
\midrule
\multicolumn{4}{c}{With Ekman pumping} \\
\texttt{LSB} & $1.5231\times 10^7$ & $14$ & $-1.2705\times 10^4$ \\
\cite{Gillet07} & $1.53\times 10^7$ & $14$ & $-1.268\times 10^4$ \\  
\bottomrule
\end{tabular}
\label{tab:onset_Gillet}
\end{table}

\begin{figure*}
 \centering
 \includegraphics[width=16cm]{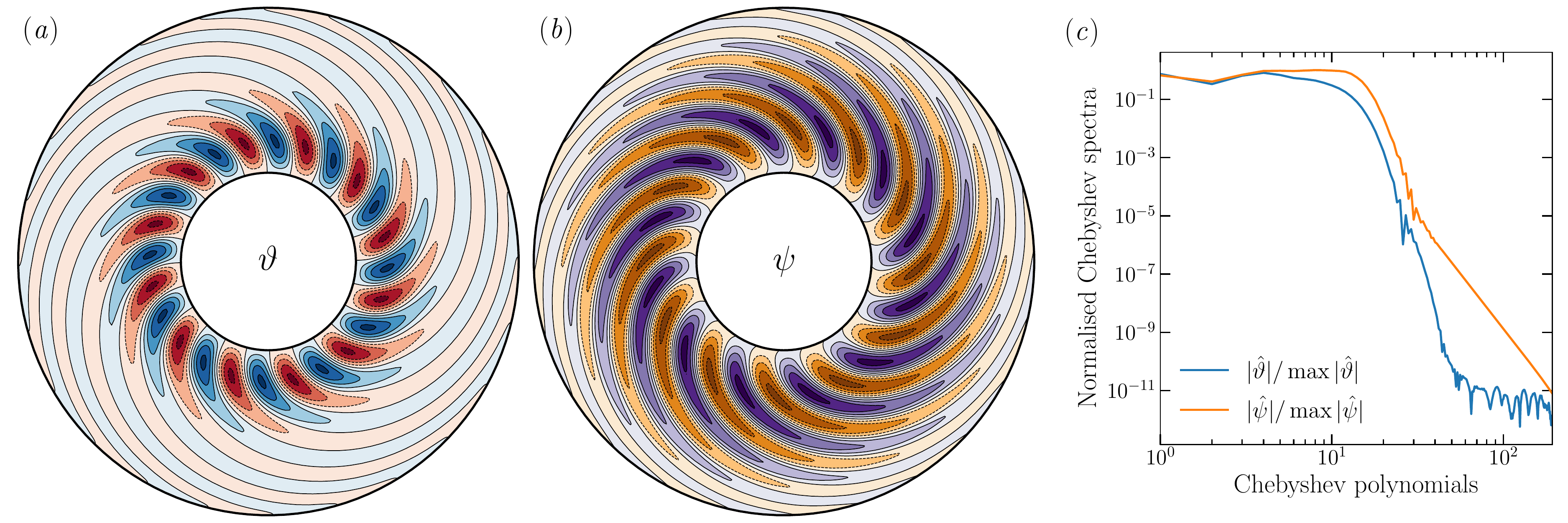}
 \caption{Eigenfunction of the first unstable mode for $E=3\times 10^{-6}$ 
and $Pr=0.025$. This mode has a critical Rayleigh number $Ra_c=9.55263\times 
10^{6}$, a drift frequency $\omega_d=-9.42690\times 10^{3}$ 
and an azimuthal wavenumber $m=12$. (\textit{a}) Temperature fluctuation 
$\vartheta$ in the equatorial plane. (\textit{b}) Streamfunction $\psi$ in the 
equatorial plane. (\textit{c}) Normalised Chebyshev spectra of the 
eigenfunction.}
 \label{fig:eigenvalue}
\end{figure*}

In the following we focus on weakly nonlinear QG convection with $E=3\times 
10^{-6}$ and $Pr=0.025$ and a radius ratio $r_i/r_o=0.35$, a physical set up 
that is quite similar to the one considered by \cite{Gillet07} for liquid 
Gallium.
Figure~\ref{fig:eigenvalue} shows the critical eigenmode (with $\tau \simeq 0$) 
computed with \texttt{LSB} for these parameters. The onset of convection takes 
the form of a thermal Rossby wave that drifts in the retrograde 
direction with a critical azimuthal wavenumber $m=12$, a drifting frequency 
$\omega_d=-9.42690\times 10^{3}$ and a critical Rayleigh number 
$Ra_c=9.55263\times 10^{6}$. The numerical convergence of this 
calculation has been assessed by computing the Chebyshev spectra of the 
different eigenfunctions as illustrated on Fig.~\ref{fig:eigenvalue}c.

\begin{figure*}
 \centering
 \includegraphics[width=16cm]{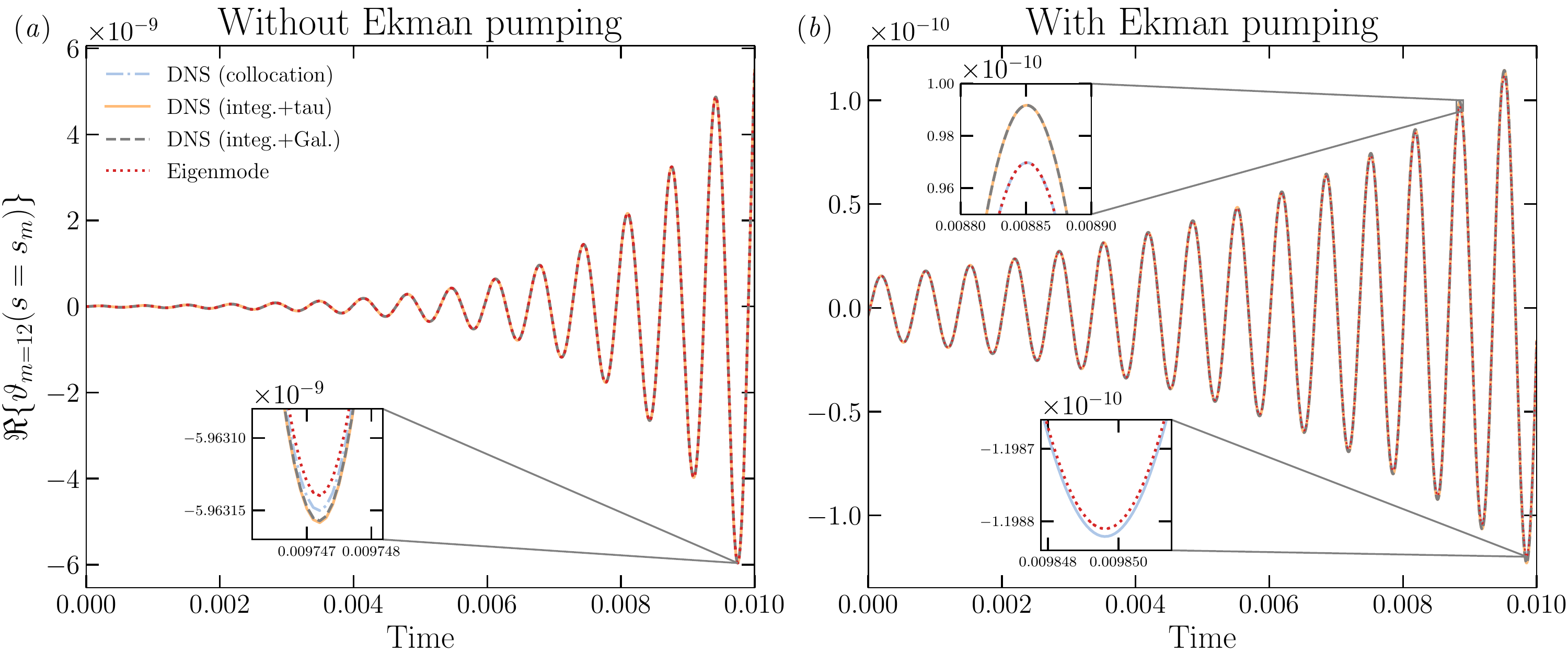}
 \caption{Real part of $\vartheta_{m=12}$ at mid-depth $s_m=0.5(s_i+s_o)$ as a 
function of time without Ekman pumping (left panel) and with Ekman pumping 
(right panel) for a case with $E=3\times 10^{-6}$, $Ra=10^7$ and $Pr=0.025$. 
Zoomed-in insets highlight the differences between the eigenmode 
and the three spatial discretisation strategies implemented in \texttt{pizza}. 
The DNS have been time-advanced using the BPR353 time scheme with a fixed time 
step size $\delta t=10^{-7}$ to ensure that the error of the time scheme is 
negligible (see Fig.~\ref{fig:error_DeltaT}). The simulations have been 
initiated with the most unstable $m=12$ eigenmode calculated with 
\texttt{LSB}. Both sparse Chebyshev formulations assume 
$\epsilon=3\times 10^{-3}$ for the cases with Ekman pumping.} 
 \label{fig:compLinDNS}
\end{figure*}

To validate the numerical 
implementation, the growth rate and the drift frequency obtained with 
\texttt{pizza} are compared to the eigenvalues derived with \texttt{LSB}. 
This requires a finite growth rate $\tau$, hence we adopt in the 
following a marginally supercritical Rayleigh number $Ra=10^7$ and compute the 
most critical eigenmodes for this $Ra$ both in absence and in presence of Ekman 
pumping. The corresponding eigenmodes $(\psi,\vartheta)^T$ computed with 
\texttt{LSB} are then used as starting conditions in \texttt{pizza}. A 
meaningful comparison 
necessitates that the nonlinear calculation remains in the weakly 
nonlinear regime. We hence restrict the computation to a short time interval of 
$10^{-2}$ viscous time, which roughly corresponds to 15 periods of the most 
unstable drifting thermal Rossby wave. To ensure that the numerical error is 
dominated by the spatial discretisation rather than by the temporal one, we 
employ the BPR353 time scheme with a small time step size $\delta t=10^{-7}$ 
(see Fig.~\ref{fig:error_DeltaT}). Figure~\ref{fig:compLinDNS} shows a 
comparison of the time evolution of the temperature fluctuation 
$\Re(\vartheta_{m=12})$ at mid depth using the linear eigenmode calculated with 
\texttt{LSB} and using the different radial discretisation schemes implemented 
in \texttt{pizza}. In absence of Ekman pumping (left panels), the different 
radial schemes yield almost indiscernible time evolution curves. The zoomed-in 
inset reveals a 6 significant digits agreement between the eigenmode and the 
weakly nonlinear calculations. When the Ekman pumping contribution is included 
(right panels), similar accuracy is recovered between the simulation
computed with the collocation method and the eigenmode. 
The two nonlinear calculations that use the Chebyshev integration approach show 
a more pronounced deviation due to the approximated Ekman pumping term with 
$\epsilon=3\times 10^{-3}$.

\begin{table*}
 \caption{Growth rate and drift frequency for the $m=12$ mode for $E=3\times 
10^{-6}$, $Ra=10^{7}$ and $Pr=0.025$ with and without Ekman pumping. The first 
line has been computed with the linear solver \texttt{LSB}, while the others 
correspond to nonlinear calculations performed with \texttt{pizza}. All the 
simulations have been computed with a fixed time step size 
$\delta t=10^{-7}$. The setups highlighted with an asterisk have been 
time-advanced with an explicit treatment of the buoyancy term. The correct 
digits compared to the eigenmode are underlined for each solution.}
 \centering
 \begin{tabular}{lcccccc}
  \toprule
             &&&\multicolumn{2}{c}{Without Ekman pumping} &
             \multicolumn{2}{c}{With Ekman pumping} \\
  $t$ scheme & $(N_r,N_c,N_m)$ & $\epsilon$  &  $\tau$ & $\omega_d$ & $\tau$ & 
$\omega_d$ \\
	 \midrule
   \multicolumn{7}{c}{Eigensolver \texttt{LSB}} \\
  - & $(192,192,1)$ & - &$6.149994\times 10^2$  & $-9.536952\times 10^{3}$ & 
$2.122883\times 10^2$ & $-9.436506\times 10^{3}$\\
  \midrule
\multicolumn{7}{c}{Chebyshev collocation} \\
  CNAB2 & (193,193,128) & - & $\underline{6.1}50091\times10^{2}$ & 
$\underline{-9.53695}1\times10^{3}$ & $\underline{2.12}3007\times10^{2}$ & 
$\underline{-9.436506}\times10^{3}$ \\
BPR353 & (193,193,128) & - & $\underline{6.14999}6\times10^{2}$ & 
$\underline{-9.53695}3\times10^{3}$ & $\underline{2.1228}92\times10^{2}$ & 
$\underline{-9.436506}\times10^{3}$ \\
SBDF3 & (193,193,128) & - & $\underline{6.1}50048\times10^{2}$ & 
$\underline{-9.53695}3\times10^{3}$ & $\underline{2.122}955\times10^{2}$ & 
$\underline{-9.436506}\times10^{3}$ \\
SBDF4 & (193,193,128) & - & $\underline{6.1}50092\times10^{2}$ & 
$\underline{-9.536952}\times10^{3}$ & $\underline{2.12}3010\times10^{2}$ & 
$\underline{-9.43650}7\times10^{3}$ \\
\midrule
\multicolumn{7}{c}{Chebyshev integration + Galerkin} \\
CNAB2 & (193,128,128) & $3\times 10^{-3}$ & $\underline{6.1}50015\times10^{2}$ & 
$\underline{-9.536952}\times10^{3}$ & $\underline{2.1}48132\times10^{2}$ & 
$\underline{-9.436}744\times10^{3}$ \\
CNAB2 & (768,512,128) & $10^{-4}$ & $\underline{6.1}50015\times10^{2}$ & 
$\underline{-9.536952}\times10^{3}$ & $\underline{2.12}3818\times10^{2}$ & 
$\underline{-9.4365}12\times10^{3}$ \\
BPR353* & (193,128,128) & $3\times 10^{-3}$ & 
$\underline{6.14999}7\times10^{2}$ & $\underline{-9.53695}3\times10^{3}$ & 
$\underline{2.1}48114\times10^{2}$ & $\underline{-9.436}745\times10^{3}$ \\
SBDF3 & (193,128,128) & $3\times 10^{-3}$ & $\underline{6.14999}7\times10^{2}$ & 
$\underline{-9.53695}3\times10^{3}$ & $\underline{2.1}48114\times10^{2}$ & 
$\underline{-9.436}745\times10^{3}$ \\
\midrule
  \multicolumn{7}{c}{Chebyshev integration + tau-Lanczos} \\
  BPR353* & (193,128,128) & $3\times 10^{-3}$ & 
$\underline{6.14999}8\times10^{2}$ & $\underline{-9.53695}3\times10^{3}$ & 
$\underline{2.1}48113\times10^{2}$ & $\underline{-9.436}745\times10^{3}$ \\
BPR353* & (769,512,128) & $10^{-4}$ & $\underline{6.14999}5\times10^{2}$ & 
$\underline{-9.53695}3\times10^{3}$ & $\underline{2.12}3799\times10^{2}$ & 
$\underline{-9.4365}13\times10^{3}$ \\
  BPR353* & (3073,2048,128) & $10^{-5}$ & $\underline{6.14999}6\times10^{2}$ & 
$\underline{-9.53695}3\times10^{3}$ & $\underline{2.122}983\times10^{2}$ & 
$\underline{-9.43650}7\times10^{3}$ \\
  \bottomrule
 \end{tabular}
 \label{tab:growth_rates}
\end{table*}

To determine the growth rate and the drift frequency in the nonlinear 
calculations, we fit the time evolution of
$\Re(\vartheta_{m=12})$ at mid depth with the function $a_0 \cos(\omega_d t 
+\phi_0) e^{\tau t}$ using least squares, the initial amplitude $a_0$ and 
phase shift $\phi_0$ being determined by the starting conditions. 
Table~\ref{tab:growth_rates} shows the obtained eigenpairs for the different 
radial schemes tested with several time integrators and values of 
$\epsilon$.
Overall the best agreement with the eigenvalues are obtained when the 
third-order BPR353 time scheme is employed. The superiority of the
SDIRK scheme likely has to do with the lack of self-starting capabilities 
of multistep schemes, which hence require a lower-order starting time stepper 
to complete the first iterations. This procedure introduces errors larger than 
the theoretical order of the scheme that could account for the slightly larger 
inaccuracy of those schemes. The approximation of the Ekman pumping 
contribution when the Chebyshev integration method is used introduces an error 
that is more pronounced in the growth rate than in the drift frequency. This is 
expected since dissipation processes usually have a direct impact on the 
growth rate of an instability. A decrease of $\epsilon$ goes 
along with a proportional drop of the relative error on $\tau$. This is however 
accompanied by an increase of the number of radial grid points in order to 
maintain the spectral convergence of the Ekman pumping term 
(\ref{eq:approx_pump}).

This comparison validates the implementation of all the linear terms that enter 
Eqs.~(\ref{eq:psi}-\ref{eq:temp}) for the different radial discretisation 
schemes. The approximation of the Ekman pumping contribution yields 
relative error that grow with $\epsilon$. The collocation method 
should hence be privileged for small problem size. Because of its fastest 
execution time, the sparse Chebyshev formulation is the 
recommended approach when dealing with larger problem sizes. A large 
number of radial grid points indeed permits to accommodate small values of 
$\epsilon < 10^{-3}$, for which the error associated with the approximate Ekman 
pumping term becomes negligible.

\subsection{Nonlinear convection}

\begin{figure*}
 \centering
 \includegraphics[width=\textwidth]{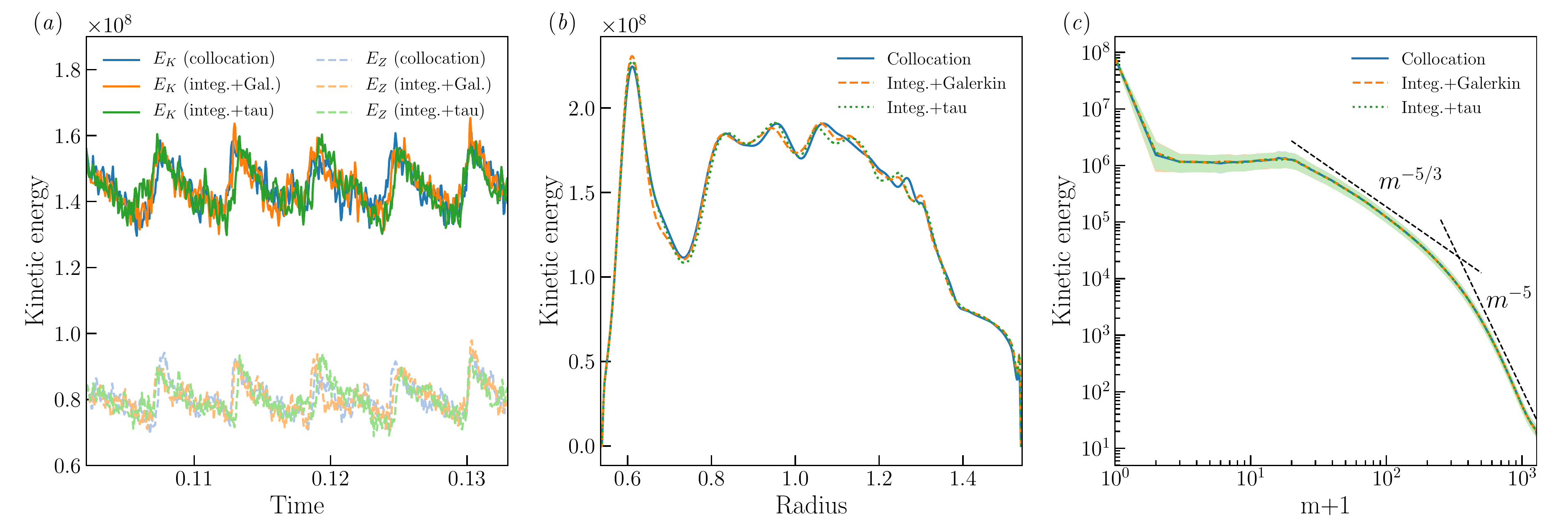}
 \caption{(\textit{a}) Total $E_K$ and zonal $E_Z$ kinetic energy
as a function of time for three numerical simulations with different radial  
discretisation schemes. (\textit{b}) Time and 
azimuthally averaged kinetic energy as a function of radius. (\textit{c}) 
Time-averaged kinetic energy spectra as a function of the wavenumber $m$. The 
shaded region correspond to one standard deviation
of temporal fluctuations relative to the time averages.
The simulations assume the following control parameters: 
$E=10^{-7}$, $Ra=2\times 10^{11}$ and $Pr=1$. The sparse cases have been 
computed with $\epsilon=10^{-3}$.}
 \label{fig:compE1e7}
\end{figure*}

\begin{table*}
  \caption{Time-averaged diagnostics of three numerical simulations with 
$E=10^{-7}$, $Ra=2\times 10^{11}$ and $Pr=1$. The simulations have been 
computed with the BPR353 time scheme. The fourth 
column corresponds to the average time step size. The fifth and sixth column 
contain the time-average and the standard deviation of the total and the zonal 
kinetic energy, respectively. The last column corresponds to the total 
number of core hours spent to compute the time interval displayed in 
Fig.~\ref{fig:compE1e7}.}
 \centering
\begin{tabular}{lccccccc}
\toprule
$r$ scheme & $(N_r,N_c,N_m)$ & $\epsilon$ & $\overline{\delta t}$ & 
$\overline{E_K}\pm\sigma(E_K)$ & $\overline{E_Z}\pm\sigma(E_Z)$ & Core hours \\
\midrule
Collocation & $(641,641,1280)$  &-  & $2.623\times 10^{-8}$ & $1.448\times 
10^{8}\pm6.281\times 10^{6}$ & $8.075\times 10^{7}\pm4.855\times 10^{6}$ & 
$1.8\times 10^{4}$ \\
Integ.+Galerkin & $(1025,682,1280)$ & $10^{-3}$ & $2.052\times 10^{-8}$ & 
$1.448\times 10^{8}\pm6.381\times 10^{6}$ & $8.038\times 10^{7}\pm4.884\times 
10^{6}$ & $3.7\times 10^{3}$ \\
Integ.+tau & $(1025,682,1280)$ & $10^{-3}$ & $2.095\times 10^{-8}$ & 
$1.443\times 10^{8}\pm6.439\times 10^{6}$ & $8.026\times 10^{7}\pm4.661\times 
10^{6}$, & $3.6\times 10^{3}$ \\
Integ.+Galerkin & $(1025,682,1280)$ & $10^{-4}$ & $2.100\times 10^{-8}$ & 
$1.452\times 
10^{8}\pm5.754\times 10^{6}$ & $8.012\times 10^{7}\pm4.211\times 10^{6}$ & 
$3.8\times 10^{3}$ \\
Integ.+tau & $(1025,682,1280)$ & $10^{-4}$ & $2.144\times 
10^{-8}$ & $1.450\times 10^{8}\pm6.663\times 10^{6}$ & $8.070\times 
10^{7}\pm5.231\times 10^{6}$ & $3.6\times 10^{3}$ \\
\bottomrule
\end{tabular}
 \label{tab:E1e7}
\end{table*}

To pursue the code validation procedure, we now examine another physical setup 
which is not in the weakly nonlinear regime anymore with $E=10^{-7}$, $Pr=1$ 
and $Ra=2\times 10^{11}$, roughly 60 times the critical Rayleigh number.
This corresponds to the setup that has been previously used to determine 
the Courant number of the different time schemes in 
\S~\ref{sec:tschemes}. To compare the different radial 
discretisation schemes, we first compute a simulation until a statistically 
steady-state has been reached. We then use this physical solution as a 
starting condition of several numerical simulations that use different 
radial discretisation schemes and two values of $\epsilon$ with the BPR353 time 
scheme. Since this is now a 
turbulent convection model, the time step size will change over time to satisfy 
the Courant condition (Eq.~\ref{eq:cfl}). To avoid the costly reconstruction of 
the matrices at each iteration, we adopt a time step size that is three quarter 
of the maximum eligible time step. The simulations are then computed over a 
timespan of roughly $0.03$ viscous time, which corresponds to more than $150$ 
turnover times. 

Figure~\ref{fig:compE1e7}a shows the time evolution of the total and the zonal 
kinetic energy defined by
\[
 E_K = \dfrac{1}{2}\left\langle u_s^2 + u_\phi^2 \right\rangle =
  E_Z+2\pi\sum_{m=1}^{N_m} \int_{s_i}^{s_o} 
\left(|u_s^m|^2+|u_\phi^m|^2\right) s\,	\mathrm{d}s\,,
\]
where the zonal contribution is expressed by
\[
 E_Z = \dfrac{1}{2}\left\langle \overline{u_\phi}^2 \right\rangle =
 \pi \int_{s_i}^{s_0} \overline{u_\phi}^2 s\,\mathrm{d} s\,.
\]
The three numerical simulations feature a very similar time evolution with 
roughly 50\% of the energy content in the axisymmetric azimuthal motions.
They show a quasi-periodic behaviour with quick energy increases followed by 
slower relaxations. This can be attributed to the time evolution of the 
zonal jets that slowly drift towards the inner boundary where they become 
unstable \citep{Rotvig07}. Panels b and c of Fig.~\ref{fig:compE1e7} show the 
time-average radial profiles and $m$ spectra of the kinetic energy, 
respectively. A good agreement is found between the three radial 
discretisation schemes. Typical of 2-D QG turbulence, an inverse energy cascade 
with a $m^{-5/3}$ slope takes place up to a typical lengthscale where the 
convective features are sheared apart by the zonal jets 
\citep[here $m\simeq 20$, see][]{Rhines75}.  At smaller lengthscales the 
spectra transition to a $m^{-5}$ slope frequently observed in Rossby waves 
turbulence \citep[e.g.][]{Rhines75,Schaeffer05a}.

For a better quantification of the difference between the three radial schemes, 
Tab.~\ref{tab:E1e7} contains the time-average and the standard deviation of 
$E_K$ and $E_Z$ over the entire run time. Since dealiasing is also required in 
the radial direction when using a sparse Chebyshev formulation, the two cases 
that have been computed with the Chebyshev integration method have a larger 
number of radial grid points to ensure a number of Chebyshev modes comparable 
to 
the one used with the collocation method. Because of the change of the grid 
spacing (Eq.~\ref{eq:cfl}), this implies a decrease in the average time-step 
size. The time averages and standard deviation obtained for the three schemes 
and the two values of $\epsilon$ are found to agree within less than 1\%. Given 
the unsteady nature of the 
solution, the differences in time step size and the limited time span 
considered for time averaging, it is not clear whether this difference 
can solely be attributed to the parametrisation of the Ekman pumping 
contribution. Notwithstanding this possible source of error, this comparison 
demonstrates that turbulent convection can be accurately modelled by an 
efficient sparse Chebyshev formulation with an acceptable error introduced by 
the Ekman pumping term approximation.

\subsection{Turbulent QG convection}

\begin{figure*}
 \centering
 \includegraphics[width=17.5cm]{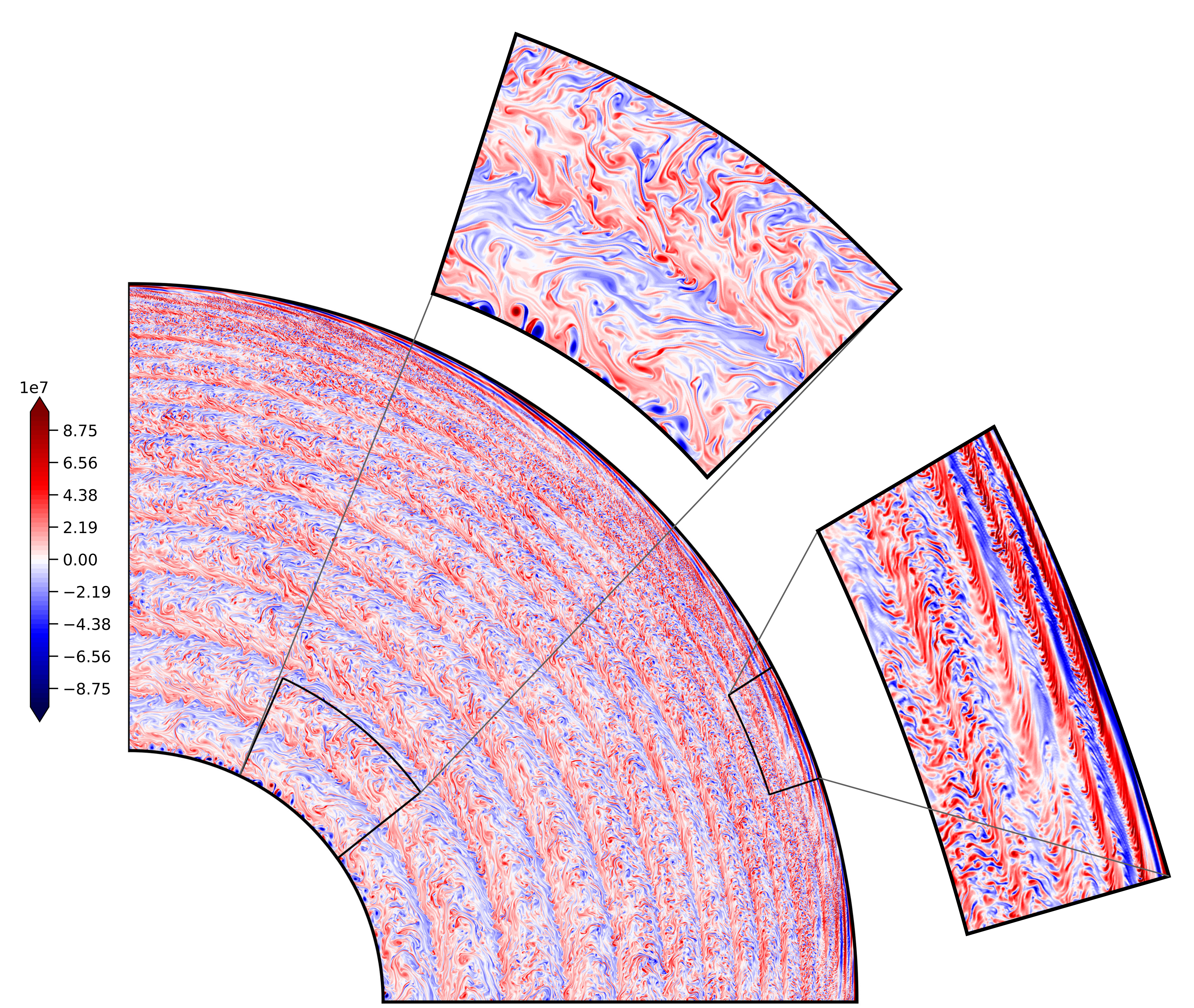}
 \caption{Snapshot of the axial vorticity for a numerical simulation 
with $E=10^{-9}$, $Ra=1.5\times 10^{14}$ and $Pr=1$. The
Chebyshev integration formulation with a Galerkin approach to enforce the 
boundary condition has been employed to compute this numerical model.
We use a spatial resolution $(N_r,N_m)=(6145,6144)$ and $\epsilon=10^{-4}$ for 
the approximated Ekman pumping term. For a better visualisation, only one 
quarter of the solution is displayed.}
 \label{fig:E1e9}
\end{figure*}

To check the ability of the spectral radial discretisation schemes to model 
turbulent QG convection, we consider a third numerical configuration with 
$E=10^{-9}$, $Ra=1.5\times 10^{14}$ and $Pr=1$. This corresponds to
strongly supercritical convection ($Ra > 100\,Ra_c$) at a very low Ekman 
number, a prerequisite to ensure that both large Reynolds and small Rossby 
numbers are reached at the same time. With the dimensionless units adopted in 
this study,
\[
 Re= \left[\dfrac{2 E_K}{\pi(s_o^2-s_i^2)}\right]^{1/2},\ Ro=Re\,E\,.
\]
For these control parameters, convection develops in the so-called turbulent QG 
regime \citep[e.g.][]{Julien12a} with $Re\simeq 10^5$ and $Ro\simeq 10^{-4}$.
Numerical models that operate at these extreme parameters demand a large 
number of grid points --here $(N_r,N_m)=(6145,6144)$-- which becomes 
intractable for the Chebyshev collocation 
method. We hence only compute this model using the Chebyshev integration method 
combined with a Galerkin approach to enforce the boundary conditions.
For this physical configuration, a time integration of roughly ten 
convective overturns requires about $10^5$ core hours.

Figure~\ref{fig:E1e9} shows a snapshot of the vorticity with two zoomed-in 
insets that emphasise the regions close the boundaries. The mixing 
of the potential vorticity $(\omega +2/E)/h$ by turbulent convective motions 
generates multiple zonal jets with alternated directions 
\citep[e.g.][]{Dritschel08}. This gives rise to a spatial separation of the 
vortical structures with alternated concentric rings of cyclonic ($\omega > 0$) 
and anticyclonic ($\omega < 0$) vorticity. The typical size of these zonal jets 
is usually well-predicted by the Rhines scale defined by 
$(Ro/|\beta|)^{1/2}$ 
\citep[e.g.][]{Rhines75,Gastine14,Verhoeven14,Heimpel16,Guervilly17}. This 
lengthscale marks the separation between Rossby waves at larger scales and 
turbulent motions at smaller scales. Because of the increase of $|\beta|$ with 
the cylindrical radius $s$ in spherical geometry, the zonal jets are getting 
thinner outward. Close to the outer boundary, the dynamics becomes dominated by 
tilted vortices elongated in the azimuthal direction, a typical pattern
of the propagation of thermal Rossby waves. Because of the steepening of 
$\beta$ at large radii, the vortex stretching term becomes the dominant source 
of vorticity there, such that the propagation of thermal Rossby waves takes 
over the nonlinear advective processes. This outer region is hence expected to 
shrink with an increase of the convective forcing \citep[e.g.][]{Guervilly17}. 
At the interface between jets, the vortical structures are sheared apart into 
elongated filaments, indicating a direct cascade of enstrophy towards 
smaller scales.

\section{Conclusion}

\label{sec:conclusion}

In this study, we have presented a new open-source code, 
nicknamed \texttt{pizza}, dedicated to the study of rapidly-rotating convection 
under the 2-D spherical quasi-geostrophic approximation 
\citep[e.g.][]{Busse86,Aubert03,Gillet06}. The code is available at 
\url{https://github.com/magic-sph/pizza} as a free software that can be used, 
modified, and redistributed under the terms of the GNU GPL v3 license. The 
radial discretisation relies on a decomposition in Fourier series in the 
azimuthal direction and in Chebyshev polynomials in the radial direction. For 
the latter, both a classical Chebyshev collocation method 
\citep[e.g.][]{Glatzmaier84,Boyd01} and a sparse integration method 
\citep[e.g.][]{Stellmach08,Muite10,Marti16} are supported.
We adopt a pseudo-spectral approach where the nonlinear advective terms are 
treated in the physical space and transformed to the spectral space using fast
discrete Fourier and Chebyshev transforms. \texttt{pizza} supports several 
implicit-explicit time schemes encompassing multi-step schemes as well as 
diagonally-implicit Runge-Kutta schemes \citep[e.g.][]{Ascher97} that have 
been validated by convergence tests. The parallelisation strategy relies  on a 
message-passing communication framework based on the \texttt{MPI} standard. The 
code has been tested and validated against onset of quasi-geostrophic 
convection. 

The comparison of the 
two radial discretisation schemes has revealed the superiority of the Chebyshev 
integration method. In contrast to the collocation technique that requires the 
storage and the inversion of dense matrices, the integration method indeed only 
involves sparse operators. As a consequence, the memory requirements only grows 
with $\mathcal{O}(N_r)$ and the operation count with $\mathcal{O}(N_r\ln N_r)$ 
as compared to $\mathcal{O}(N_r^2)$ when using a 
collocation approach. Multi-step and diagonally-implicit Runge-Kutta schemes 
have shown comparable efficiency, defined in this study by the ratio of the 
maximum CFL number over the numerical cost of one iteration. Additional 
parameter studies with various Reynolds and Rossby numbers are however required 
to assess the differences between both families of time integrators.
We have found a good parallel scaling up to 
roughly four radial grid points per \texttt{MPI} task. This implies that large 
spatial resolution up to $\mathcal{O}(10^4\times 10^4)$ grid points can be 
reached with a reasonable wall time if one uses several thousands of 
\texttt{MPI} tasks. Such large grid resolutions allows the study of 
turbulent quasi-geostrophic convection at low Ekman numbers.
Preliminary results for a numerical model with 
$E=10^{-9}$, $Ra=1.5\times 10^{14}$ and $Pr=1$ shows the formation of multiple 
zonal jets, when both the Reynolds number is large $\mathcal{O}(10^5)$ and the 
Rossby number is small $\mathcal{O}(10^{-4})$. This specific combination of $Re 
\gg 1$ and $Ro \ll 1$ is a prerequisite to study the turbulent quasi-geostrophic 
convection regime \citep{Julien12a}, an important milestone to better 
understand the internal dynamics of planetary interiors.
%
%
%

Future developments of the code include the implementation of
the time-evolution of chemical composition to study double-diffusive 
convection under the spherical QG framework. On the longer term, the QG flow 
and temperature computed in the equatorial plane of the spherical shell will be 
coupled to an induction equation computed in the entire shell using 
classical 3-D pseudo-spectral discretisation \citep[e.g.][]{Schaeffer06}.

\begin{acknowledgments}
I want to thank Alexandre Fournier for his comments that helped to improve the 
manuscript. Stephan Stellmach and Benjamin Miquel are acknowledged for their 
fruitful advices about Galerkin bases and Philippe Marti for his help with the 
symbolic \texttt{python} package used to assemble the sparse Chebyshev matrices. 
I also wish to thank Michel Rieutord for sharing the \texttt{Linear Solver 
Builder} eigensolver. Numerical computations have been carried out on the 
S-CAPAD platform at IPGP and on the \texttt{occigen} cluster at GENCI-CINES 
(Grant A0020410095). All the figures have been generated using 
\texttt{matplotlib} \citep{Hunter07}. All the post-processing tools that have 
been used to construct the different figures are part of the source code of 
\texttt{pizza} and are hence freely accessible. This is IPGP contribution 4015.
\end{acknowledgments}

\bibliographystyle{gji}

\appendix

\section{Direct solve of a bordered matrix}
\label{sec:app1}

Suppose one wants to solve the following linear problem which involves a 
so-called bordered matrix $\mathcal{A}$
\[
 \mathcal{A} \psi = f,
\]
where $\mathcal{A}$ comprises $p$ full top rows and a
banded structure underneath. The matrix problem is sub-divided as follows
\[
 \left(
   \begin{array}{cc}
A & B\\  C & D
\end{array}\right) \left( \begin{array}{c} \psi_1  \\ \psi_2 \end{array}  
\right) =
\left(\begin{array}{c} g \\ h \end{array}\right),
\]
where $A$ is a full square matrix of size $(p\times p)$, $B$ is a full matrix 
of size $(p\times n-p)$, $C$ is a sparse matrix of size $(n-p\times p)$ and $D$ 
is a band matrix of size $(n-p)$ with a bandwidth $q$, $q$ being the total 
number of bands.	
One first solves the two following banded linear problems
\[
 D x = h\,, \quad D y = C\,.
\]
The LU factorisation of the band matrix $D$ requires $\mathcal{O}(q^2\,n)$ 
operations, while the solve  requires $\mathcal{O}(q\,n)$ 
operations \citep[e.g.][Appendix~B2]{Boyd01}. We then assemble the Schur 
complement of the banded block $D$
\[
 M = A - B D^{-1} C = A -By\,,
\]
before solving the small dense problem of size $(p,p)$ 
\[
 M \psi_1 = g-Bx\,.
\]
This requires $\mathcal{O}(p^3)$ operations for the LU factorisation and 
$\mathcal{O}(p^2)$ for the solve. This cost remains negligible as long as $p 
\ll n$, which is the case for the linear problems considered in the Chebyshev 
integration method. We finally evaluate
\[
 \psi_2 = x - y\,\psi_1\,,
\]
to assemble the final solution given by $\psi=(\psi_1,\psi_2)^T$.

\section{Galerkin basis for streamfunction boundary conditions}
\label{sec:app2}

In this section, we derive a Galerkin basis function for the following 
combination of boundary conditions that is used in the Chebyshev integration 
method for the streamfunction equation
\[
 \varPsi = \dfrac{\partial \varPsi}{\partial 
s}= 0, \quad\text{for}\quad s=s_i\,,
\]
and
\[
 \varPsi = \dfrac{\partial^3 \varPsi}{\partial 
s^3}= 0, \quad\text{for}\quad s=s_o\,.
\]
We start by defining the following ansatz for the Galerkin set
\[
 \phi_n(x) = \sum_{i=0}^{4} \gamma_{i}^n\, T_{n+i}(x)\,.
\]
Following \cite{McFadden90} and \cite{Julien09} we then make use of the tau 
boundary conditions 
(Eqs.~\ref{eq:bcs_coll_dirichlet},\ref{eq:bcs_coll_neumann} and 
\ref{eq:bc_d3psi_coll}) to form the following system of equations
\[
  \begin{aligned}
 \phi_n(1) & = \sum_{i=0}^4 \gamma_i^n &=0, \\
 \phi_n(-1) & = \sum_{i=0}^4 (-1)^{i} \gamma_i^n &=0 , \\ 
  \dfrac{\partial^3\phi_n}{\partial x^3}(1)& = 
\sum_{i=0}^4 (n+i)^2[(n+i)^2-1][(n+i)^2-4] \gamma_{i}^n 
& = 0, \\  
 \dfrac{\partial\phi_n}{\partial x}(-1) &= \sum_{i=0}^{4} (-1)^{i+1} (n+i)^2 
\gamma_{i}^n &= 0,  
  \end{aligned}
\]
Since there are only four equations for five unknowns, there is a degree of 
freedom in the determination of the coefficients. We thus choose in following
\[
 \gamma_0^n = 1,
\]
which yields the following identities for the other coefficients:
\[
 \begin{aligned}
 \gamma_1^n &=\frac{8 \left(n + 1\right) \left(n^{2} + 4 n + 5\right)}{2 n^{4} 
+ 
20 n^{3} + 78 n^{2} + 140 n + 95}, \\
 \gamma_2^n &=- \frac{2 \left(n + 2\right) \left(2 n^{4} + 16 n^{3} + 58 n^{2} 
+ 
104 n + 75\right)}{\left(n + 3\right) \left(2 n^{4} + 20 n^{3} + 78 n^{2} + 140 
n + 95\right)}, \\
 \gamma_3^n &=- \frac{8 \left(n + 1\right) \left(n^{2} + 4 n + 5\right)}{2 
n^{4} 
+ 20 n^{3} + 78 n^{2} + 140 n + 95},
\end{aligned}
\]
 and
\[
\gamma_4^n= \frac{\left(n + 1\right) \left(2 n^{4} + 12 n^{3} + 30 n^{2} + 36 n 
+ 15\right)}{\left(n + 3\right) \left(2 n^{4} + 20 n^{3} + 78 n^{2} + 140 n + 
95\right)}\,.
\]

\bsp 

\label{lastpage}

\end{document}